\documentclass[a4paper,12pt]{article}
\usepackage[dvips]{graphics}
\usepackage{epsfig}
\usepackage{psfrag}
\usepackage[psamsfonts]{amssymb}
\usepackage{amsmath}
\usepackage{indentfirst}
\usepackage{amssymb}
\usepackage{wrapfig}
\newcommand{\be}{\begin{equation}}
\newcommand{\ee}{\end{equation}}
\newcommand{\ba}{\begin{eqnarray}}
\newcommand{\ea}{\end{eqnarray}}

\begin{document}

\centerline{\Large\bf Theory of Earthquake Recurrence Times}

\vskip 1cm \centerline{A. Saichev$^1$ and D. Sornette$^{2,3}$}
\vskip0.5cm

\centerline{$^1$ Mathematical Department, Nizhny Novgorod State
University} \centerline{Gagarin prosp. 23, Nizhny Novgorod, 603950,
Russia}

\centerline{$^2$ D-MTEC, ETH Zurich, CH-8032 Z\"urich, Switzerland
(email: dsornette@ethz.ch)}

\centerline{$^3$ LPMC, CNRS UMR6622 and Universit\'e des Sciences,
Parc Valrose, 06108 Nice Cedex 2, France}

\vskip 1cm

{\bf Abstract}: The statistics of recurrence times in broad areas
have been reported to obey universal scaling laws, both for single
homogeneous regions (Corral, 2003) and when averaged over multiple
regions (Bak et al.,2002). These unified scaling laws are
characterized by intermediate  power law asymptotics.
On the other hand, Molchan (2005) has presented a mathematical proof
that, if such a universal law exists, it is necessarily an
exponential, in obvious contradiction with the data. First, we
generalize Molchan's argument to show that an approximate unified
law can be found which is compatible with the empirical observations
when incorporating the impact of the Omori law of earthquake
triggering. We then develop the full theory of the statistics of
inter-event times in the framework of the ETAS model of triggered
seismicity and show that the empirical observations can be fully
explained. Our theoretical expression fits well the empirical
statistics over the whole range of recurrence times, accounting for
different regimes by using only the physics of triggering quantified
by Omori's law. The description of the statistics of recurrence
times over multiple regions requires an additional subtle
statistical derivation that maps the fractal geometry of earthquake
epicenters onto the distribution of the average seismic rates in
multiple regions. This yields a prediction in excellent agreement
with the empirical data for reasonable values of the fractal
dimension $d \approx 1.8$, the average clustering ratio $n \approx
0.9$, and the productivity exponent $\alpha \approx 0.9$ times the
$b$-value of the Gutenberg-Richter law. Our predictions are remarkably
robust with respect to the magnitude threshold used to select observable
events. These results extend previous works which have shown that much
of the empirical phenomenology of seismicity can be explained by
carefully taking into account the physics of triggering between
earthquakes.

\pagebreak

\section{Introduction}

The concept of ``recurrence time'' (also called recurrence interval, inter-event time
or return time) is widely used for hazard assessment in seismology.
The average recurrence time of an earthquake is usually defined as
the number of years between occurrences of an earthquake of a given
magnitude in a particular area. Once chosen a probabilistic model
for the distribution of recurrence times (often a Poisson
distribution in earlier implementations, evolving progressively to
more realistic distributions), an estimation of the probability for
damaging earthquakes in a given area over a specific time horizon
can be derived. For instance, the Working Group on Califonia
Earthquake Probabilities (2003) assessed a 0.62 probability of a
major damaging earthquake striking the greater San Francisco Bay
Region over the following 30 years (2002-2031). Such calculations
are based on classifications of known active faults and on
assumptions about the organization of seismicity on these faults,
guided by geological, paleoseismic (see for instance Sieh, 1981),
geodetic evidence (Global Earthquake Satellite System, 2003) as well
as earthquake patterns extracted from recent seismic catalogs.
Starting with the simple assumption that faults are independent and
carry their own characteristic earthquake (Schwartz and Coppersmith,
1984), seismologists and other scientists working on seismic hazards
are realizing that more complex models are needed to take into
account the interaction, coupling and competition between faults
(Lee et al., 1999) which translates into a rich phenomenology in the
space-time-magnitude organization of earthquakes.

A new view is now emerging that recurrence times should be
considered for broad areas, rather than for individual faults, and
could provide important insights in the physical mechanisms of
earthquakes. For instance, Wyss et al. (2000) have proposed to
relate an indirect measurement of local recurrence times within
faults to geometrical asperities and stress field, thus
significantly broadening the standard notion of return time.
Inspired by the scaling approach developed in the physics of
critical phenomena (Sornette, 2004), Kossobokov and Mazhkenov
(1988), Bak et al. (2002) and Christensen et al. (2002) have
proposed a unified scaling law combining the Gutenberg-Richter law,
the Omori law and the fractal distribution of epicenters to describe
the distributions of inter-event times between successive
earthquakes in a hierarchy of spatial domain sizes and magnitudes in
Southern California. Corral (2003, 2004a,b, 2005a) has refined and
extended these analyses to many different regions of the world and
has proposed the existence of a universal scaling law for the
probability density function (PDF) $H(\tau)$ of recurrence times (or
inter-event times) $\tau$ between earthquakes in a given region $S$:
\begin{equation}
H(\tau)\simeq \lambda f(\lambda \tau)~ . \label{1}
\end{equation}
The remarkable proposition is that the function $f(x)$, which
exhibit different power law regimes with cross-overs, is found
almost the same for many different seismic regions, suggesting
universality. The specificity of a given region seems to be captured
solely by the average rate $\lambda$ of observable events in that
region, which fixes the only relevant characteristic time
$1/\lambda$ for the recurrence times.

The interpretation proposed by Bak et al. (2002), Christensen et al.
(2002) and Corral (2003, 2004a,b, 2005a) is that the scaling law
(\ref{1}) reveals a complex spatio-temporal organization of
seismicity, which can be viewed as an intermittent flow of energy
released within a self-organized (critical?) system, for which
concepts and tools from the theory of critical phenomena can be
applied (Corral, 2005b). This view has been challenged by Lindman et
al. (2005) who stressed several methodological caveats (see Corral
and Christensen (2006) for a reply). Livina et al. (2005) have
additionally noticed that the marginal (or mono-variate) PDF of
inter-event times gives only a partial description of the time
sequence of earthquakes in a given region, as it is found to be a
function of preceding inter-event times. There is thus a memory
between successive earthquakes which influences the distribution of
the inter-event times, a conclusion which is well-known to most
seismologists.

It is fair to state that there is at present no theoretical
understanding of these empirical results and in particular of
expression (\ref{1}). The situation becomes even more interesting
with the recent mathematical demonstration by Molchan (2005) that,
under very weak and general conditions, the only possible form for
$f(x)$, if universality holds, is the exponential function, in
strong disagreement with the observations reported by Bak et al.
(2002) and Corral (2003, 2004a,b, 2005a). In addition, from a
re-analysis of the seismicity of Southern California, Molchan and
Kronrod (2005) have shown that the unified scaling law (\ref{1}) is
incompatible with multifractality which seems to offer a better
description of the data.

Here, our purpose is to show how all the above can be simply
understood and reconciled from the standard known statistical laws
of seismicity:
\begin{itemize}
\item{(i)} the Gutenberg-Richter distribution $\sim
1/E^{1+\beta}$ (with $\beta \approx (2/3)b \approx 2/3$) of
earthquake energies $E$ (Knopoff et al., 1982);

\item{(ii)} the Omori law $\sim 1/t^p$ (with $p \approx 1$ for
large earthquakes) of the rate of aftershocks as a function of time
$t$ since a mainshock (Utsu et al., 1995);

\item{(iii)} the productivity law $\sim
E^{a}$ (with $a \approx 2/3$) giving the number of earthquakes
triggered by an event of energy $E$ (Helmstetter et al., 2005);

\item{(iv)} the fractal (and even probably multifractal
(Ouillon et al., 1996)) structure of fault networks (Davy et al,
1990) and of the set of earthquake epicenters (Kagan and Knopoff,
1980).
\end{itemize}
The question we address can be summarized as follows: are the
statistics on inter-event times described in (Bak et al., 2002;
Corral, 2003, 2004a,b, 2005a; Livina et al., 2005) really new in the
sense that they reveal some information which is not contained in
the above laws (i-iv), as claimed by these authors? Or can they be
derived from the known statistical properties of seismicity, so that
they are only different ways of presenting the same information?

In order to address this question, we use the simplest possible
model which combines these four laws, framed as a stochastic point
process of past earthquakes triggering future earthquakes: the Omori
law (ii) is taken to describe the conditional rate of activation of
new earthquakes, given all past earthquakes; the Gutenberg-Richter
law (i) describes the distribution according to which the magnitude
of a new earthquake is independently determined; the productivity
law (iii) gives the weight of the contribution of a given past
earthquake in the production of new earthquakes. In order to take
into account the fractal geometry of earthquake catalogs, the new
earthquakes can be positioned on a fractal geometry. The elements
(i-iii) actually constitute the basic building block of the
benchmark model of seismicity, known as the Epidemic-Type Aftershock
Sequence (ETAS) model of triggered seismicity (Ogata, 1988; Kagan
and Knopoff, 1981), whose main statistical properties are reviewed
in (Helmstetter and Sornette, 2002). Various versions have been or
are currently applied by different groups to describe observed and
forecast future seismicity (Console et al., 2002; 2003a,b; 2006;
Console and Murru, 2001; Gerstenberger et al., 2005; Reasenberg, P.
A. and Jones, 1989; 1994; Steacy et al., 2005).

The common characteristics of this class of models is to treat all
earthquakes on the same footing such that one does not assume any
distinction between foreshocks, mainshocks and aftershocks: each
earthquake is considered to be capable of triggering other
earthquakes according to the three basic laws (i-iii) mentioned
above. This hypothesis is based on the realization that there are no
observable differences between foreshocks, mainshocks and
aftershocks (Jones et al., 1999; Helmstetter and Sornette, 2003a;
Helmstetter et al., 2003), notwithstanding their usual
classification based in retrospect analysis of realized seismic
sequences: Earthquakes preceding the mainshock are called foreshocks
and earthquakes following the mainshock are called aftershocks but
no physical differences between these earthquakes are known. For
instance, a mainshock (classified as the largest event in a given
space-time window) may be re-classified as a foreshock if a larger
event later follows it; and what would have been an aftershock
becomes a mainshock if larger than all its preceding earthquakes.

In this paper, we use specifically the ETAS model in the version
proposed by Ogata (1988). The ETAS model assumes that earthquakes
magnitudes are mutually statistically independent and drawn from the
Gutenberg-Richter (GR) probability $Q(m)$. The GR law gives the
probability that magnitudes of triggered events are larger than a
given level $m$ (the relationship between magnitude $m$ and energy
is $m \propto (2/3) \ln_{10} E$). We shall use the GR law in the
form
\begin{equation}
Q(m_2)= Q(m_1)10^{-b(m_2-m_1)}~ , \label{GR}
\end{equation}
which emphasizes its scale invariance property. Here $b \simeq 1$
and $m_1$, $m_2$ are arbitrary magnitudes. We parameterize the
(bare) Omori law (Sornette and Sornette, 1999; Helmstetter and
Sornette, 2002) for the rate of triggered events of first-generation
from a given earthquake as
\begin{equation}
\Phi(t)= {\theta c^\theta \over (c+t)^{1+\theta}} \label{Omori}
\end{equation}
with $\theta \gtrsim 0$. One may interpret $\Phi(t)$ as the PDF of
random times of independently occurring first-generation
aftershocks, triggered by some mainshock which happened at the
origin $t=0$. The last ingredient is the productivity law, which we
write in the following convenient form for further analysis,
\begin{equation}
\rho(m)= \kappa ~10^{\alpha(m-m_0)}~ , \label{productivity}
\end{equation}
where the factor $\kappa$ will be related below to physically
observable quantities such as the average branching ratio $n$ (which
is the average number of triggered earthquakes of first generation
per triggering event). The magnitude $m_0$ is a cut-off introduced
to regularize the theory (Helmstetter and Sornette, 2002). It can be
interpreted as the smallest possible magnitude for earthquakes to be
able to trigger other earthquakes (Sornette and Werner, 2005a).
Several authors have shown that the ETAS model provides a good
description of many of the regularities of seismicity (Console et
al., 2002; 2003a,b; 2006; Console and Murru, 2001; Helmstetter and
Sornette, 2003a,b; Helmstetter et al., 2005; Gerstenberger et al.,
2005; Ogata, 1988; 2005; Ogata and Zhuang, 2006; Reasenberg, P. A.
and Jones, 1989; 1994; Saichev and Sornette, 2005; 2006a; Steacy et
al., 2005; Zhuang et al., 2002; 2004; 2005).

Our main result is that, according to Occam's razor, the previously
mentioned results on universal scaling laws of inter-event times do
not reveal more information than what is already captured by the
well-known laws (i-iii) of seismicity (Gutenberg-Richter, Omori,
essentially), together with the assumption that all earthquakes are
similar (no distinction between foreshocks, mainshocks and
aftershocks), which is the key ingredient of the ETAS model. Our
theory is able to account quantitatively for the empirical power
laws found by Bak et al. and Corral, showing that they result from
subtle cross-overs rather than being genuine asymptotic scaling
laws. We also show that universality does not strictly hold.

The organization of the paper is the following. In section~2, we
discuss in detail Molchan's derivation (Molchan, 2005) that, if the
scaling law (\ref{1}) holds true, then it should be exponential in
the sense that $f(x)=e^{-x}$. We extend Molchan's argument by
developing a simple semi-quantitative theory, showing that the
presence of the Omori law destroys the self-similarity of the PDF of
recurrence times. Nevertheless, if the exponent of the Omori law
(\ref{Omori}) is close to $1$, i.e., $\theta \ll 1$, then, the PDF
of recurrence times approximately obeys a non-exponential scaling
law which can fit well the empirical data. Section~3 extends further
the discussion of section ~2 by proposing simplified models of
aftershock triggering process, in the goal of demonstrating the
direct relation between the Omori law and the corresponding scaling
law for the PDF of the recurrence times for an arbitrary region.
These discussions allow us to stress the generality of the curve that
we propose to the disagreement between Molchan's result and
empirical data. They also prepare us to understand better the full
derivation using the technology of generating probability functions
(GPF) applied to the ETAS model. In section~4, we analyze the ETAS
model of triggered seismicity with the formalism of GPF and
establish the main general results useful for the following. As a
check for the formalism, we obtain the statistical description for
the number of observable events which occur within a given region
and during the time window $[t,t+\tau]$. Section 5 exploit the
general results of section~4 to obtain predictions on the PDF of
recurrence times between observable events in a single homogeneous
region characterized by a well-defined average seismic rate. A short
account of some of these results is presented in (Saichev and
Sornette, 2006c). Section~6 combines the previous results on single
regions to obtain predictions of the PDF of inter-event times
between earthquakes averaged over many different regions. In a first
part, we construct different ad hoc models of the distribution of
the average seismic rates of different regions. In a second part, we
propose a procedure converting the fractal geometry of epicenters
into a specific distribution of the average seismic rates of
different regions. The knowledge of this distribution allows us to
compute the PDF of inter-event times averaged over many regions
which is compared with Bak et al. (2002)'s and Corral (2004a)'s
empirical analyses. We find a very good agreement between our
prediction and the empirical PDF of inter-event times.

\section{Generalized Molchan's relation}

We start by addressing the puzzle raised by the demonstration by
Molchan (2005) based on a probabilistic reasoning that, under very
weak and general conditions, the only possible form for $f(x)$ in
(\ref{1}), if universality holds, is the exponential function, in
contradiction with the strongly non-exponential form of the reported
seemingly universal unified PDF (\ref{1}) for inter-event times.
Here, we briefly reproduce Molchan's argumentation and then
generalize it to show that the presence of the Omori law, while
destroying the exact unified law, gives nevertheless an approximate
unified law fitting rather precisely the real data.

Let $P(\tau)$ be the probability that there are no observable events
during the time interval $[t,t+\tau]$ within some region $S$. Let us
call $\lambda$ the average rate of observable events within the
region $S$. Let us assume the existence of a unified law
\begin{equation}
P(\tau)= \varphi(\lambda\tau)~ , \label{unified}
\end{equation}
valid for any region. In expression (\ref{unified}), $\varphi(x)$ is
the universal scaling function, which is assumed to be the same for
all regions. Following Molchan's argument, if an unified law
(\ref{unified}) holds true for any region, then it should be valid
for the region made of the union of two disjoint regions
characterized respectively by the rates $\lambda_1$ and $\lambda_2$.
If the triggering processes of earthquake in those two regions are
statistically independent, then the following functional equation
should be true
\begin{equation}
\varphi((\lambda_1+\lambda_2)\tau)=\varphi(\lambda_1 \tau)
\varphi(\lambda_2 \tau)~ . \label{Molchan}
\end{equation}
Posing $\psi(x)= \ln\varphi(x)$, the equation (\ref{Molchan}) is
equivalent to
\begin{equation}
\psi(x)= \psi\left( {\lambda_1 \over \lambda_1 + \lambda_2} x
\right)+ \psi\left( {\lambda_2 \over \lambda_1 + \lambda_2} x
\right)~ . \label{Molchan linear}
\end{equation}
It is well-known that the solution of equation (\ref{Molchan
linear}) defines the class of linear funtions $\psi(x)=-a x$ for
arbitrary $a$'s, leading to the exponential form \be \varphi(x)=
e^{-ax}~. \label{mbnmbsa} \ee

For the PDF (\ref{unified}), there is a normalization condition that
allows us to fix the constant $a$. We need first to recall a few
general results of the theory of point processes (Daley and
Vere-Jones, 1988) Let the random times \be \dots t_{-1}, t_0, t_1,
\dots t_k \dots \qquad t_{k+1}>t_k>t_{k-1} \dots \ee form a
stationary point process with an average rate equal to $\lambda$.
Then, the PDF $H(\tau)$ of the inter-event times $t_{k+1}-t_k$ can
be obtained from the following relation
\begin{equation}
H(\tau)= {1 \over \lambda} {d^2 P(\tau) \over d\tau^2}~,
\label{d2pdf}
\end{equation}
where $P(\tau)$ is the already mentioned probability of absence of
events within the interval $[t,t+\tau]$. The normalization condition
for $H(\tau)$ reads \be \int_0^\infty H(\tau)d\tau= -{1\over
\lambda} {d P(\tau) \over d\tau} \Big|_{\tau=0}=
\varphi'(x)\big|_{x=0}= 1~ . \ee Substituting with (\ref{mbnmbsa})
yields $a=1$. Thus, following Molchan's reasoning, if a unified law
(\ref{1}) for the PDF of recurrence times exists, then it should be
exponential \be f(x)= \varphi''(x)= e^{-x}~ . \label{mhmbmvs} \ee

We have already mentioned that the law (\ref{mhmbmvs}) derived by
Molchan (2005) contradicts the observations (Bak et al., 2002;
Corral, 2003, 2004a,b, 2005a). Let us thus generalize Molchan's
reasoning, by assuming that the probability $P(\tau)$ depends, not
only on the dimensionless combination $\lambda \tau$, but
additionally on the average rate $\lambda$. This means that we
assume that the probability $P(\tau)$ can expressed as
\begin{equation}
P(\tau)=\varphi(\lambda \tau,\lambda)~ , \label{11}
\end{equation}
for some universal function $\varphi$. With this assumption,
equation (\ref{Molchan}) is replaced by \be
\varphi((\lambda_1+\lambda_2)\tau,\lambda_1+\lambda_2)=\varphi(\lambda_1
\tau,\lambda_1) \varphi(\lambda_2 \tau,\lambda_2)~, \ee and
expression (\ref{Molchan linear}) is changed into
\begin{equation}
\psi(x,\lambda_1+\lambda_2)= \psi\left( {\lambda_1 \over \lambda_1 +
\lambda_2} x, \lambda_1 \right)+ \psi\left( {\lambda_2 \over
\lambda_1 + \lambda_2} x, \lambda_2 \right)~ ,\label{linear}
\end{equation}
where again $\psi= \ln\varphi$ and $x$ is a dimensionless
variable. It is easy to check that the solution of (\ref{linear})
has the form \be \psi(x,\lambda)= - x ~g\left({x \over
\lambda}\right)~ , \label{mmkafq} \ee where $g(y)$ is a function
which is arbitrary except that it should be such that
$\psi(x,\lambda)$ is monotonically decreasing with respect to $x$.
Expression (\ref{mmkafq}) translates into
\begin{equation}
P(\tau) = e^{- \lambda \tau g(\tau)}~ . \label{nonexp}
\end{equation}
It is clear from expression (\ref{nonexp}) that $g(\tau)$ should be
dimensionless, hence its argument must be dimensionless. Since $g$
is independent of $\lambda$, the only possibility to make $g(\tau)$
dimensionless is that it depends on another time scale $c$ such that
$g$ can be written. \be g(\tau)= g_0\left({\tau \over c}\right)~ ,
\ee where $g_0(y)$ is a function with dimensionless argument. This
implies that expression (\ref{nonexp}) has actually the form
\begin{equation}
P(\tau) = \exp\left(- \lambda \tau g_0\left({\tau \over c}\right)
\right)~ , \label{nonexpevent}
\end{equation}
which generalizes (\ref{mbnmbsa}). Note that the later is recovered
for the special case where the function $g_0$ is a constant (which
has then to be unity by the condition of normalization discussed
above).

The relevance of a time scale $c$ in addition to the inverse rate
$1/\lambda$ is actually part of Omori's law. Consider the often used
pure power form of the Omori law \be \Phi(t) = k~ t^{-1-\theta}~ .
\ee Then, necessarily, for $\theta \neq 0$, a time scale $c$ is
needed so that $\Phi$ has the dimension of the inverse of time,
i.e., \be k = k_0 c^\theta \ee where $k_0$ is a dimensionless
constant.

It appears natural on physical grounds (and will be justified in our
calculations below) that the Omori law has an influence on the form
of the probability $P(\tau)$. This suggests that the physical origin
of the time scale $c$ (and hence of the deviation
(\ref{nonexpevent}) from Molchan's law (\ref{mbnmbsa})) lies in
Omori's law. In other words, this argues for the fact that the
function $g_0\left({\tau \over c}\right)$ in (\ref{nonexpevent})
actually derives from Omori's law. We thus obtain the key insight
that the existence of the Omori law implies the absence of a
universal scaling law for the PDF of inter-event times!

In our calculations presented below using the ETAS model, we will
show that the Omori law gives the following specific structure for
the function $g_0(y)$: \be g_0(y)= a + h y^{-\theta}~. \ee
Substituting it into (\ref{nonexpevent}) and using the dimensionless
variable $x=\lambda \tau$, we obtain
\begin{equation}
P(\tau)= \varphi(x,\epsilon)=\exp\left(-ax - \epsilon^\theta h
x^{1-\theta}\right)~ , \qquad \epsilon=\lambda c~ . \label{pmod}
\end{equation}
A remarkable property of the distribution $\varphi(x,\epsilon)$ is
that it changes very slowly for $\theta\ll 1$, even if the average
rate $\lambda$ changes by factors of thousands. This characteristic
property of $\varphi(x,\epsilon)$ is at the origin, as we show
below, of the essentially non-exponential approximate unified law
for the PDF (\ref{1}) of recurrence times, which provides excellent
fit to the PDF's of real data.

\section{Simplified model of the impact of Omori's law on the PDF of inter-event times}

In the preceding section, we have suggested that the Omori law may
be at the origin of both (i) a breaking of the unified PDF of the
inter-event times from the exponential function derived by Molchan
and (ii) an approximate universal law different from the
exponential. In the following sections, we will present a rigorous
analysis of the PDF of recurrence times in the framework of the ETAS
model, which confirms this claim. In the present section, we provide
what we believe is a simple intuitive understanding of the roots of
the observed approximate unified law, by using a simplified model of
recurrence time statistics which takes into account the impact of
the Omori law.

We consider a synthetic Earth in which two types of earthquakes
occur. Spontaneous events are triggered by the driven tectonic
forces. In turn, these spontaneous events (also called
``immigrants'' in the jargon of branching processes) may trigger
their ``aftershocks'' (called more generally ``triggered events'').
We put quotation marks around ``aftershocks'' to stress the fact
that those ``aftershocks'' may be larger than their mother event.
They are not necessarily the aftershocks of the nomenclature of
standard seismology. These triggered events of first generation may
themselves be the sources of triggered events of the second
generation and so on. We denote $\omega$ the Poisson rate of the
observable spontaneous events. The average number of
first-generation ``aftershocks,'' triggered by some spontaneous
event, defines the key parameter $n$ of the theory, often called the
branching parameter. For the process to be stationary and not
explode, we consider the sub-critical regime $n<1$ (Helmstetter and
Sornette, 2002). Note that $n$ is by definition the average number
of first-generation events triggered by any arbitrary earthquake. It
has also the physical meaning of being the fraction of triggered
events in a catalogue including both spontaneous sources and
triggered events over all possible generations (Helmstetter and
Sornette, 2003c). Given the average rate $\omega$ of the spontaneous
sources, the average rate of all events, including spontaneous and
all their children over all generations, is $\omega (1+ n +n^2 +
...)$, that is,
\begin{equation}
\lambda= {\omega \over 1-n}~ . \label{meanrate}
\end{equation}
In reality, one does not have the luxury of observing all
earthquakes. Catalogues are complete only above a minimum magnitude
which depends on the density of spatial coverage of the network of
seismic stations and on their quality. One thus typically observes
only a small subset of all occurring earthquakes, with the vast
majority (of small earthquakes) being hidden from detection. Here,
we formulate a simplifying hypothesis, justified later by a full
rigorous calculation with the ETAS model, that, due to the
statistical independence of the magnitudes of earthquakes, the ratio
of the number of observable spontaneous events to the number of all
observable events is approximately independent of the magnitude
threshold $m$. Assuming this to be true, then the relation
(\ref{meanrate}) also holds between the rate $\omega(m)$ of
observable spontaneous events and the rate $\lambda(m)$ of the total
set of observable events, whose magnitudes all exceed the threshold
level $m$.

We take into account the influence of both the spontaneous sources
and of the triggered events on the probability $P(\tau)$ of abscence
of events within time interval $[t,t+\tau]$, by postulating the
following generalized Poisson statistics
\begin{equation}
P(\tau) \approx \exp\left(-\omega \tau- \omega \Lambda(\tau)
\right)~ . \label{pimpact}
\end{equation}
The first term $\omega\tau$ in the exponential is nothing but the
standard Poisson contribution that no spontaneous events fall inside
the time interval $[t,t+\tau]$. The second term $\omega
\Lambda(\tau)$ describes the contribution resulting from all the
events which have been triggered before $t$. Neglecting for
simplicity the difference of productivity of different events, we
use the simplified form
\begin{equation}
\omega \Lambda(\tau)=\int_t^{t+\tau} \lambda(t') dt'~ , \qquad
\lambda(t')\approx \sum_{t_k<t} \Phi(t'-t_k)~ , \label{meanlambda}
\end{equation}
where $\Phi(t)$ is the normalized Omori law described by relation
(\ref{Omori}). We assume that the Omori law is self-averaging, so
that we can replace $\lambda(t')$ by its average value \be
\langle\lambda\rangle(t'-t)= {\omega n \over 1-n} \int_{-\infty}^t
\Phi(t'-t'') dt''= {\omega n \over 1-n} a(t'-t)~ . \ee Here, $\omega
n/(1-n)$ is the average number per unit time of triggered events
occurring before the time window $[t,t+\tau]$ and
\begin{equation}
a(t)= \int_t^\infty \Phi(t') dt' = {c^\theta \over (c+t)^\theta}~ .
\label{at}
\end{equation}

Replacing in (\ref{meanlambda}) $\lambda(t')$ by
$\langle\lambda\rangle(t'-t)$ yields \be \Lambda(\tau)\approx { n
\over 1-n} A(\tau)~ , \ee where
\begin{equation}
A(\tau)= \int_0^\tau a(t)dt= {c \over 1-\theta} \left[ \left(1+{\tau
\over c}\right)^{1-\theta}-1\right]~ . \label{alarge}
\end{equation}
In particular, if $\tau\gg c$, we have \be \Lambda(\tau)\approx
{\omega  n \over 1-n} {c\over 1-\theta} \left({\tau \over
c}\right)^{1-\theta}~ , \qquad \tau \gg c~ . \ee Thus, the
simplified model (\ref{pimpact}) for the probability that no events
occur in the time interval $[t,t+\tau]$ takes the form
\begin{equation}
P(\tau)= \exp\left(- \omega \tau - {\omega n \over 1-n} A(\tau)
\right)~ , \label{psimp}
\end{equation}
which, for $\tau\gg c$, reads
\begin{equation}
P(\tau)\approx \exp\left(-\omega \tau- {\omega n  \over 1-n} {c\over
1-\theta} \left({\tau \over c}\right)^{1-\theta}\right)~ , \quad
\tau\gg c~ . \label{psimptgc}
\end{equation}
Using relation (\ref{meanrate}) and introducing the dimensionless
parameters
\begin{equation}
x= \lambda \tau~ , \qquad \epsilon= \lambda c~ , \label{apho}
\end{equation}
we obtain finally
\begin{equation}
P(\tau) \approx \varphi(x,\epsilon)=\exp\left(- (1-n) x -{ n
\epsilon^\theta \over 1-\theta} x^{1-\theta} \right)~ .
\label{pimpactx}
\end{equation}
This expression (\ref{pimpactx}) coincides with the previously
proposed form (\ref{pmod}) with the correspondence $a=1-n$ and
$h=n/(1-\theta)$.

By substituting (\ref{pimpactx}) into (\ref{d2pdf}), we obtain that
$H(\tau)$ is described by the relation (\ref{1}), where
\begin{equation}
\displaystyle f(x)= \left(n \epsilon^\theta \theta x^{-1-\theta}+
\left[1-n+ n \epsilon^\theta x^{-\theta}\right]^2\right)
\varphi(x,\epsilon) ~ . \label{pdfrho}
\end{equation}

\begin{figure}[h]
\centerline{\includegraphics[width=14cm]{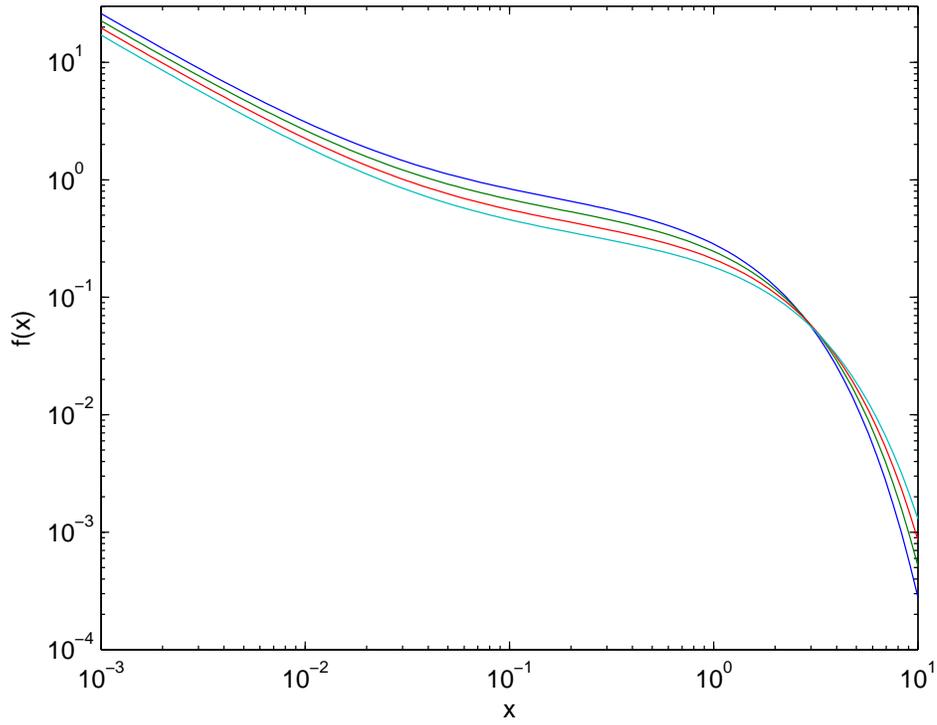}}
\caption{\label{Fig1} Plots of the PDF $f(x)$ (\ref{pdfrho}) for
$n=0.9$, $\theta=0.03$, for different threshold magnitudes $m-m_1=
0;2;4;6$ (top to bottom on the left), where the reference parameter
is $\epsilon_1=\lambda_1 c=10^{-4}$ for $m_1$. }
\end{figure}

The PDF $f(x)$ (\ref{pdfrho}) depends on the scale $L$ of the region
$S$ under observation, via the dependence of the parameter
$\epsilon$ on the rate $\lambda$ which itself depends on $L$.
Similarly, via the same chain of dependence, $f(x)$ is also a
function of the threshold magnitude $m$ of observed events. However,
the PDF $f(x)$ given by (\ref{pdfrho}) is very slowly varying with
the average rate $\lambda$ for $\theta\ll 1$, which is typical of
aftershock sequences. When $\lambda$ varies by large factors up to
thousands, $\epsilon^\theta$ changes insignificantly. Indeed, let us
denote by $\lambda_1$ and $\lambda_2$ the average rates
corresponding to two magnitude threshold levels $m_1$ and $m_2$.
Using the GR law, the parameter $\epsilon^\theta$ changes by the
factor $\rho$ times ($\epsilon_2^\theta= \rho \epsilon_1^\theta$)
where \be \rho= \left({\lambda_2 \over \lambda_1}\right)^\theta=
10^{-b(m_2-m_1)}~. \ee Consider for instance $m_2-m_1=4$, for which
the average rate changes by the huge factor $10^{-4}$ while the
$\rho\simeq 0.76$ still remains rather close to $1$ for
$\theta=0.03$. Fig.~\ref{Fig1} plots the PDF $f(x)$ given by
(\ref{pdfrho}) for $n=0.9$, $\theta=0.03$ and for different
magnitude thresholds $m$. This figure demonstrates the slow
dependence of $f(x)$ with respect to magnitude threshold level $m$.

\begin{figure}[h]
\centerline{\includegraphics[width=14cm]{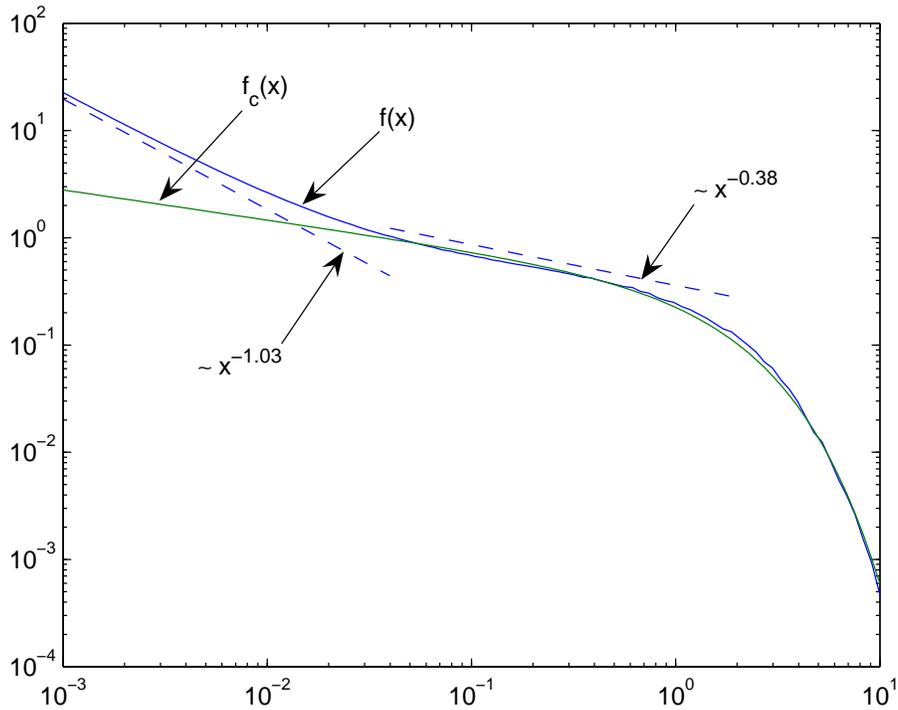}}
\caption{\label{Fig2} Plot of the PDF $f(x)$ given by expression
(\ref{pdfrho}) for the parameters $n=0.9$, $\theta=0.03$,
$\epsilon_1=0.76$, $m-m_1= 2$, and Corral's fitting curve
(\ref{Corralfit}) for $\gamma=0.38$, $\delta=1$, $d=1.7$ and
$C=0.75$. We also show an intermediate asymptotics $\sim 1/x^{0.38}$
discussed by Corral (see below) as well as the short-time
asymptotics $\sim 1/x^{1.03}$ corresponding to the Omori law. }
\end{figure}

While the plots of the approximate scaling law (\ref{pdfrho}) shown
in Fig.~\ref{Fig1} differ significantly from Molchan's exponential
unified law, they are close to the parameterization of the empirical
PDF proposed by Corral (2004a)
\begin{equation}
f_c(x)= {C \delta \over d ~\Gamma(\gamma/\delta)} \left({x \over a}
\right)^{\gamma-1} e^{-(x/a)^\delta} \label{Corralfit}~ ,
\end{equation}
with the parameters \be \gamma= 0.67\pm 0.05~ , \qquad \delta=1.05
\pm 0.05~ , \qquad a=1.64 \pm 0.15~ , \ee and with a normalization
factor $C$ close to 1. Fig.~\ref{Fig2} shows a typical PDF $f(x)$
obtained from our expression (\ref{pdfrho}) together with Corral's
fitting curve (\ref{Corralfit}), illustrating their closeness for
not too small $x$'s. The two expressions depart for very small $x
\lesssim 0.01$, for which our simplified model distribution $f(x)$
given by (\ref{pdfrho}) exhibits a power asymptotic $f(x)\sim
x^{-1-\theta}$, which is a direct consequence of the Omori law. This
asymptotic term results explicitely from the first term in the
factor multiplying $\varphi(x,\epsilon)$ in expression
(\ref{pdfrho}). This asymptotic is absent in Corral's fitting
function $f_c(x)$ given by (\ref{Corralfit}), although it actually
exists in real data plots (Corral, 2004a). We will emphasize this
point later when we discuss the full theory and its comparison with
the empirical data.

\section{Description of the statistics of observable events
in the context of the ETAS model}

In the preceding sections, we have made plausible and intuitively
appealing the possibility that an approximate unified law indeed
exists for the PDF of the inter-earthquake times, by using a
simplified transparent model of the statistics of recurrence times.
In this section, we develop an accurate mathematical description of
the statistical properties of observable events, in the framework of
the ETAS model presented in the introduction. The technology
developed here will then be used in the next section to quantify
precisely the prediction of the ETAS model of the PDF of inter-event
times and to derive different approximations.

Our analysis is based on our previous calculations (Saichev and
Sornette, 2006a), which showed that, for large domain sizes ($L
\sim$ tens of kilometers or more), one may neglect the impact of
aftershocks triggered by events that occurred outside the considered
spatial domain $S$, while only considering the events within the
space domain which are triggered by sources also within the same
domain. In this approximation validated by precise calculations in
(Saichev and Sornette, 2006a), the analysis of the statistical
properties of time sequence of earthquakes can be reduced to the
study of a time-only version of the ETAS model, in which the
dependence on the spatial extension of a given area can be taken
into account via the dependence on $L$ of the rate $\omega$ of
spontaneous observable events.

As the ETAS model has an exact representation in terms of branching
clusters (Hawkes and Oakes, 1974; Helmstetter and Sornette, 2002),
the main mathematical tool to analyze the ETAS model in this context
is the Generating Probability Function (GPF) of the random number
$R(t,\tau,m)$ of observable events within the time interval
$[t,t+\tau]$
\begin{equation}
\Omega(z;\tau,m)= \langle z^{R(t,\tau,m)}\rangle~ . \label{GPFdef}
\end{equation}
Here and below, the angle brackets $\langle\dots\rangle$ represent
the procedure of statistical averaging. We assume that the process
of earthquake triggering is stationary, so that the GPF given by
(\ref{GPFdef}) does not depend on $t$ but only on the duration
$\tau$ of the time interval $[t,t+\tau]$. Let $P(r,\tau,m)$ be the
probability that the number $R(t,\tau,m)$ of observable events is
equal to $r$. Then, one may rewrite the definition (\ref{GPFdef}) in
the equivalent form
\begin{equation}
\Omega(z;\tau,m)= \sum_{r=0}^\infty P(r;\tau,m) z^r~ .
\label{GPFprob}
\end{equation}
It follows that the probability that no event occurs in a time
interval of duration $\tau$ is equal to
\begin{equation}
P(\tau)= P(r=0,\tau,m)= \Omega(0;\tau,m)~ . \label{GPFzero}
\end{equation}

\begin{figure}[h]
\centerline{\includegraphics[width=14cm]{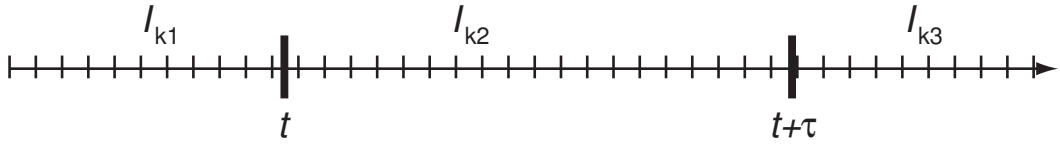}}
\caption{\label{Fig3} Partition of the time axis in small intervals
which are classified in three categories $I_{k_1}, I_{k_2}$ and
$I_{k_3}$.}
\end{figure}

To determine the GPF $\Omega(z;\tau,m_0)$ associated with the total
number of events inside the time window $[t,t+\tau]$, we proceed as
follows. We partition the $t$-axis in small intervals
$I_k=[t_k, t_k+\Delta]$ as shown in Fig.~\ref{Fig3} and then group
them in three categories: the first one $I_{k_1}\in (-\infty,t)$
lies before the time window $[t,t+\tau]$ of interest, the second one
$I_{k_2}\in [t,t+\tau]$ groups all small intervals inside the time
window $[t,t+\tau]$ and the third one $I_{k_3}\in (t+\tau,\infty)$
lies after the time window $[t,t+\tau]$. Let us now determine in
turn the impacts of the spontaneous events occurring within each of
these three intervals on the GPF of the number of events inside the
time window $[t,t+\tau]$. Let $\Theta_{k_1}(z;,m_0)$ be the GPF of
the number of spontaneous events occurring in the first class
$I_{k_1}$ of small intervals. It is equal to
\begin{equation}
\Theta_{k_1}(z;m_0)= \sum_{r=0}^\infty \mathcal{P}(r,\Delta)z^r~ ,
\label{Poissonspont}
\end{equation}
where $\mathcal{P}(r,\Delta)$ is the Poissonian probability that $r$
spontaneous events occur in $I_{k_1}$. It is given by
\begin{equation}
\mathcal{P}(r,\Delta)= {(\omega \Delta)^r \over r!} e^{-\omega
\Delta}~ .  \label{Poissonprob}
\end{equation}

Let us recall that, according to the ETAS model, spontaneous and
triggered events have statistically independent magnitudes whose PDF
is to GR law
\begin{equation}
p(m)=- {d Q(m) \over dm}= 10^{-b(m-m_0)} b \ln 10~ . \label{GRpdf}
\end{equation}
Thus, the spontaneous earthquakes have statistically independent
magnitudes $\{m_j\}$, $j=1,2,\dots,r$. According to the branching
property of the ETAS model (Hawkes and Oakes, 1974; Helmstetter and
Sornette, 2002), these spontaneous events trigger statistically
independent numbers $R_j(t_k,t,\tau,m_j,m_0)$ of events within time
interval $[t,t+\tau]$. This remark allows us to obtain the GPF
$\Omega_{k_1}(z;\tau,m_0)$ of the random number of events within
$[t,t+\tau]$ triggered by spontaneous events which occurred inside a
given small interval $I_{k_1}$: by replacing in (\ref{Poissonspont})
the term $z^r$ by \be \left[\int_{m_0}^\infty
p(m')\Theta(z;t_{k_1}-t,\tau,m',m_0)dm' \right]^r~, \ee using the
expression (\ref{Poissonprob}) and performing the summation yields
\be \Omega_{k_1}(z;\tau,m_0)= \exp\left( \omega \Delta \left[
\int_{m_0}^\infty
p(m')\Theta(z;t_{k_1}-t,\tau,m',m_0)dm'-1\right]\right)~ . \ee Here,
\be \Theta(z;t_{k_1},\tau,m',m_0)= \langle
z^{R(t_{k_1},t,\tau,m',m_0)} \rangle \ee is the GPF of the random
numbers of events triggered inside the time interval $[t,t+\tau]$ by
some spontaneous event of magnitude $m'$ which occurred at a time
$t_{k_1}<t$. Due to the statistical independence of the random
numbers of events triggered by different spontaneous events which
belong to different small intervals $I_{k_1}$, the GPF of the random
numbers of events (including all their ``aftershocks''), triggered
within the interval $[t,t+\tau]$ by spontaneous events which
occurred before the window $[t,t+\tau]$, is equal to \be
\begin{array}{c}\displaystyle
\Omega_1(z;\tau,m_0)= \prod_{k_1} \Omega_{k_1}(z;t_{k_1}\tau,m_0)= \\[3mm]
\displaystyle \exp\left( \omega \Delta \sum_{k_1}\left[
\int_{m_0}^\infty
p(m')\Theta(z;t_{k_1}-t,\tau,m',m_0)dm'-1\right]\right)~ .
\end{array}
\ee In the limit $\Delta\to 0$, the discrete sum in the exponential
becomes an integral, which yields \be \Omega_1(z;\tau,m_0)=
\exp\left( \omega \int_{-\infty}^t dt'\left[ \int_{m_0}^\infty
p(m')\Theta(z;t'-t,\tau,m',m_0)dm'-1\right]\right)~ . \ee

Analogously, one can find the GPF $\Omega_2(z;\tau,m_0)$ of the
number of windowed events (including all their ``aftershocks'')
triggered by spontaneous events which occurred within the interval
$[t,t+\tau]$. It reads \be \Omega_2(z;\tau,m_0)= \exp\left( \omega
\int_t^{t+\tau} dt'\left[ \int_{m_0}^\infty p(m')z
\Theta(z;t+\tau-t',m',m_0)dm'-1\right]\right)~ , \ee where \be
\Theta(z;\tau,m',m_0)= \Theta(z;t=0,\tau,m',m_0) \ee is the GPF of
the random number of events triggered inside the interval
$t\in[0,\tau]$ by a spontaneous event which occurred at $t=0$.

For the third contribution from spontaneous sources in $I_{k_3}\in
(t+\tau,\infty)$, the causality principle implies that
$\Theta_{k_3}\equiv 1$.

Thus, the whole GPF of the numbers of events occurring within the
time interval $[t,t+\tau]$ is $\Omega(z;\tau,m_0)=
\Omega_1(z;\tau,m_0) \Omega_2(z;\tau,m_0)$, which can be written
\begin{equation}
\Omega(z;\tau,m_0)= e^{-\omega L(z;\tau,m_0)}~ , \label{GPFall}
\end{equation}
where
\begin{equation}
\begin{array}{c}\displaystyle
L(z;\tau,m_0)= \int_{m_0}^\infty dm' p(m') \times \\[3mm]\displaystyle
\left( \int_0^\infty \left[1-\Theta(z;t,\tau,m',m_0)\right] dt
+\int_0^\tau \left[1- z \Theta(z;t,m',m_0)\right] dt\right)~ .
\end{array} \label{GPFallexp}
\end{equation}

In order to calculate GPF $\Omega(z;,\tau,m_0)$ given by
(\ref{GPFall}), (\ref{GPFallexp}), we need to determine the GPF
$\Theta(z;t,\tau,m',m_0)$ of the number of events triggered within
$[t,t+\tau]$ by some spontaneous event magnitude $m'$ which occurred
at $t'=0$. Using a similar procedure as just done for
$\Omega(z;\tau,m_0)$ and interpreting the Omori law (\ref{Omori}) as
the PDF of random times of independently triggered aftershocks, we
obtain
\begin{equation}
\begin{array}{c}
\Theta(z;t,\tau,m',m_0)= \displaystyle \exp\Big( -\rho(m')
\int_{m_0}^\infty dm'' p(m'') \times
\\[3mm]
\displaystyle  \left[b(t+\tau)-z \Phi(t+\tau)\otimes
\Theta(z;\tau,m'',m_0) -\Phi(t) \otimes \Theta(z;t,\tau,m'',m_0)
\right]\Big)~ .
\end{array} \label{GPFmainshockttau}
\end{equation}
The symbol $\otimes$ denotes the convolution operator. We also
define \be b(t)= 1-a(t)~, \ee where $a(t)$ is given by (\ref{at}).
Similarly, $\Theta(z;\tau,m',m_0)$ satisfies to the equation
\begin{equation}
\begin{array}{c}
\Theta(z;\tau,m',m_0)=\\[3mm] \exp\left( -\rho(m')
\displaystyle \int_{m_0}^\infty dm'' p(m'') \left[ b(\tau)-
z\Phi(\tau) \otimes \Theta(z;\tau,m'',m_0)\right] \right)~ .
\end{array} \label{GPFmainshocktau}
\end{equation}

The above GPF's take into account events which have arbitrary
magnitudes $m'\geqslant m_0$. In order to get the GPF (\ref{GPFdef})
which only counts observable events with magnitudes larger than some
threshold level $m$, one has to replace (i) $m_0$ by $m$ in the
r.h.s of expressions (\ref{GPFall}) and (\ref{GPFallexp}) and (ii)
$z$ by
\begin{equation}
z(m',m)= H(m- m')+z H(m'-m)~ , \label{zm}
\end{equation}
where $H(x)$ is unit step (Heaviside) function. We also replace the
GPF $\Theta(z;t,\tau,m',m_0)$ by the GPF $\Theta(z;t,\tau,m',m)$ of
the number of observable aftershocks (whose magnitudes are larger
than $m$), triggered by a mainshock of magnitude $m'$. We thus
obtain
\begin{equation}
\Omega(z;\tau,m)= e^{-\omega L(z;\tau,m)}~ , \label{GPFm}
\end{equation}
where
\begin{equation}
\begin{array}{c} \displaystyle
L(z;\tau,m)= \int_0^\infty [1- D(z;t,\tau,m)]dt+
\\[3mm] \displaystyle
\int_0^\tau [1- D(z;t,m)+ (1- z) D_+(z;t,m)] dt \label{GPFexpm}
\end{array}
\end{equation}
and
\begin{equation}
\begin{array}{c} \displaystyle
D(z;t,\tau,m)= \int_{m_0}^\infty \Theta(z;t,\tau,m',m) p(m') dm'~ ,
\\[3mm] \displaystyle
\displaystyle D(z;\tau,m)= \int_{m_0}^\infty
\Theta(z;\tau,m',m)p(m') dm'~ ,
\\[3mm] \displaystyle
\displaystyle D_+(z;\tau,m)= \int_m^\infty \Theta(z;\tau,m',m)p(m')
dm'~ .
\end{array} \label{dint}
\end{equation}
It follows from (\ref{GPFmainshockttau}) and
(\ref{GPFmainshocktau}), after replacing the GPF
$\Theta(z;,t,\tau,m',m_0)$ by $\Theta(z;,t,\tau,m',m)$, the GPF
$\Theta(z;\tau,m',m_0)$ by $\Theta(z;\tau,m',m)$ and $z$ by
$z(m'',m)$ (\ref{zm}), that the above auxiliary functions satisfy
the equations
\begin{equation}
\begin{array}{c}\displaystyle
D(z;t,\tau,m)= \Psi\big[b(t+\tau)- \Phi(t) \otimes D(z;t,\tau,m)
\\[3mm]
\displaystyle - \Phi(t+\tau) \otimes D(z;\tau,m) +(1- z)
\Phi(t+\tau) \otimes D_+(z;\tau,m) \big] \label{dttau}
\end{array}
\end{equation}
and
\begin{equation}
\begin{array}{c}
D(z;\tau,m)= \Psi\big[b(\tau)- \Phi(\tau) \otimes D(z;\tau,m) +
(1-z) \Phi(\tau) \otimes D_+(z;\tau,m)\big]~ , \\[3mm]
D_+(z;\tau,m)= \Psi_+\big[b(\tau)- \Phi(\tau) \otimes D(z;\tau,m) +
(1-z) \Phi(\tau) \otimes D_+(z;\tau,m)\big]~ . \label{dtau}
\end{array}
\end{equation}
Here,
\begin{equation}
\begin{array}{c}\displaystyle
\Psi(y)= \int_{m_0}^\infty p(m') e^{-\rho(m')y} dm'=
\gamma (\kappa y)^\gamma \Gamma(-\gamma, \kappa y)~ , \\[4mm]
\displaystyle \Psi_+(y)= \int_m^\infty p(m') e^{-\rho(m')y} dm' =
Q(m) \Psi[ Q^{-1/\gamma}(m)y] ~ . \label{psis}
\end{array}
\end{equation}
For our following calculations, it is useful to expand $\Psi(y)$
into a power series with respect to $y$:
\begin{equation}
\Psi(y)\simeq 1- n y+ \beta y^\gamma - \eta y^2 ~ ,
\label{psiseries}
\end{equation}
where \be n= {\kappa \gamma \over \gamma-1 }~ , \quad \beta=
-\left(n { \gamma-1 \over \gamma} \right)^\gamma \Gamma(1-\gamma)~ ,
\quad \eta= {n^2 (\gamma-1)^2 \over 2 \gamma (2-\gamma)}~ . \ee
Recall that $n$ is the so-called branching parameter (which we used
to obtain expression (\ref{meanrate})), defined as the average of
the total number of first-generation aftershocks triggered by some
mainshock.

Equations (\ref{dttau}) and (\ref{dtau}) are rather complicated
nonlinear integral equations. For the rest of this section, we use
them together with expressions (\ref{GPFm}), (\ref{GPFexpm}) only to
derive the average number of observable earthquakes. In the next
sections, we exploit them fully to obtain the PDF of the inter-event
times.

As recalled for instance in the Appendix of (Saichev and Sornette,
2006a), the usefulness of the theory of GPF is to obtain simple
expressions for the average number of observable earthquakes: \be
\langle R(t,\tau,m)\rangle= {d \Omega(z;\tau,m) \over
dz}\Big|_{z=1}~ . \ee Using (\ref{GPFm}), (\ref{GPFexpm}), this
leads to
\begin{equation}
\langle R(t,\tau,m)\rangle= \omega Q(m) \tau +\omega \int_0^\infty
M(t, \tau,m) dt + \omega \int_0^\tau M(\tau,m) d t~ , \label{rtau}
\end{equation}
where \be M(t,\tau,m)= {d D(z;t,\tau,m) \over dz } \Big|_{z=1}~ ,
\quad M(\tau,m)= {d D(z;\tau,m) \over dz } \Big|_{z=1}~ .
\label{nmhjhiyiy} \ee The functions in (\ref{nmhjhiyiy}) satisfy the
following equations derived from (\ref{dttau}) and (\ref{dtau}):
\begin{equation}
\begin{array}{c}
M(t,\tau,m)- n \Phi(t) \otimes M(t, \tau, m)= \\[3mm] =n \Phi(t+\tau)
\otimes M(\tau, m)+ n \left[a(t)-a(t+\tau)\right] Q(m)
\end{array} \label{mttau}
\end{equation}
and
\begin{equation}
M(\tau,m)- n \Phi(\tau) \otimes M(\tau, m)= n b(\tau) Q(m)~ .
\label{mtau}
\end{equation}

It follows from (\ref{mttau}) that the following equality is true
\be (1-n) \int_0^\infty M(t,\tau,m) dt = n a(\tau) \otimes M(\tau,m)
+ n Q(m) A(\tau)~ . \ee Using this equality in (\ref{rtau}), we
obtain \be
\begin{array}{c}
\langle R(t,\tau,m)\rangle= \\[3mm] \displaystyle \omega \left[ Q(m) \tau + {n \over 1-n} Q(m)
A(\tau) + {n \over 1-n} a(\tau) \otimes M(\tau,m) + \int_0^\tau M(t,
m) dt \right]~ ,
\end{array}
\ee where $A(\tau)$ is given by (\ref{alarge}).

Using equation (\ref{mtau}), it is easy to show that \be {n \over
1-n} a(\tau) \otimes M(\tau,m) + \int_0^\tau M(\tau, m) dt = {n
\over 1-n} Q(m) \tau -{n \over 1-n} Q(m) A(\tau)~ . \ee This leads
to
\begin{equation}
\langle R(t,\tau,m) \rangle = {\omega(m) \over 1-n}  \tau~ , \qquad
\omega(m)=\omega Q(m)~ , \label{averageobs}
\end{equation}
which has a physically intuitive interpretation: the average number
of observable events is equal to average number of observable
spontaneous events $\omega(m) \tau$ within the interval of duration
$\tau$, multiplied by the factor $1/(1-n)$ taking into account all
the triggered ``aftershocks'' of all generations. The result
(\ref{averageobs}) which can be obtained more directly (Helmstetter
and Sornette, 2002; 2003c) serves as a consistency check of our GPF
formalism.

\section{Statistics of recurrence times in the context of the ETAS model: accounting
for Corral's empirical analyses}

We now use the formalism developed in the preceding section to
derive the probability $P(\tau)$ given by (\ref{GPFzero}) that no
event occurs with a time interval of duration $\tau$. Using
(\ref{GPFm}), (\ref{GPFexpm}), (\ref{dint}), it is equal to
\begin{equation}
P(\tau)= e^{-\omega L(\tau,m)}~ , \label{pabs}
\end{equation}
where
\begin{equation}
L(\tau,m)= \int_0^\infty N(t,\tau,m) dt + \int_0^\tau N_-(t,m) dt
\label{eltauem}
\end{equation}
and
\begin{equation}
\begin{array}{c} \displaystyle
N(t,\tau,m)= 1- D(t,\tau,m)~ , \\[3mm]
\displaystyle N_-(\tau,m)=1- D(\tau,m)+D_+(\tau,m) ~ .
\end{array} \label{nns}
\end{equation}
These functions have a clear probabilistic interpretation. For
instance, $N(t,\tau,m)$ is the probability that some spontaneous
earthquake, which occurred at time $t=0$, will trigger at least one
observable event (which can be an aftershock of arbitrary
generation) within the time window $[t,t+\tau]$ ($t>0$).

It follows from (\ref{dttau}) that the probability $N(t,\tau,m)$
satisfies the equation
\begin{equation}
\begin{array}{c}
N(t,\tau,m)= \\[3mm]
1-\Psi\left[\Phi(t) \otimes N(t,\tau,m)+ \Phi(t+\tau) \otimes
N_-(\tau,m)\right]~ .
\end{array} \label{nttau}
\end{equation}
Analogously, it follows from (\ref{dtau}) that $N_-(\tau,m)$
satisfies the equation
\begin{equation}
\begin{array}{c}
N_-(\tau,m)=   \\[3mm] \displaystyle
1- \Psi\left[\Phi(\tau) \otimes N_-(\tau, m)\right] + Q(m)
\Psi\left[Q^{-1/\gamma}(m) \Phi(\tau) \otimes N_-(\tau, m)\right]~ .
\end{array} \label{ntau}
\end{equation}

\begin{figure}[h]
\centerline{\includegraphics[width=14cm]{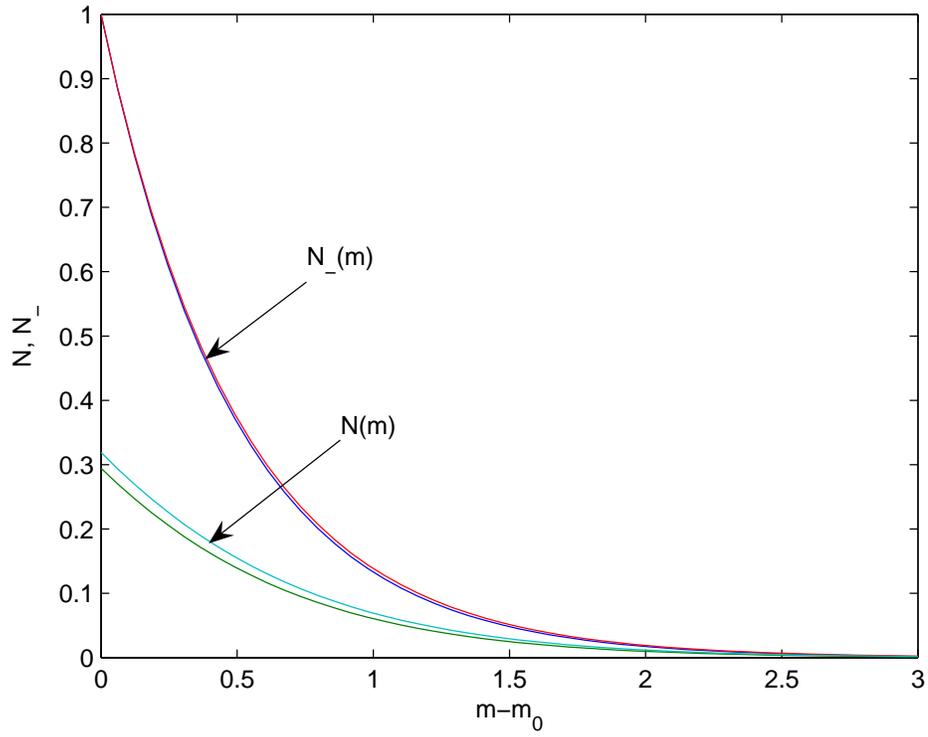}}
\caption{\label{Fig4} Plots of the solutions of the nonlinear
functional equations (\ref{eqsfornn}) for $N(m)$ and $N_-(m)$, for
the parameters $\gamma=1.2$ and two values of criticality parameter
$n=0.8; 0.9$.}
\end{figure}

These are still rather complicated nonlinear integral equations. We
can simplify them by linearization, which provides a reasonable
approximation, as can be seen from the following argument. First, we
notice that the inequalities obviously hold
\begin{equation}
N(t,\tau,m) < N(m)~ , \qquad N_-(\tau,m) < N_-(m)~ , \label{ineqnn}
\end{equation}
where \be N(m)= N(t=0,\tau=\infty,m)~ , \qquad N_-(m)=
N_-(\tau=\infty,m)~ . \ee The majorants $N(m)$ and $N_-(m)$ satisfy
the following nonlinear functional equations derived from from
(\ref{nttau}), (\ref{ntau}):
\begin{equation}
\begin{array}{c}
N(m)=1 - \Psi[N_-(m)]~ , \\[3mm]
N_-(m)=1-\Psi[N_-(m)]+ Q(m) \Psi[Q^{-1/\gamma}(m) N_-(m)]~ .
\end{array} \label{eqsfornn}
\end{equation}
We have solved numerically these equations (\ref{eqsfornn}) and
Fig.~\ref{Fig4} plots these solutions for $N(m)$ and $N_-(m)$. One
can observe that, even for moderate magnitude thresholdss
$m-m_0\gtrsim 1$, both $N(m)\ll 1$ and $N_-(m)\ll 1$, so that, with
the inequalities (\ref{ineqnn}), one can be confident that a linear
approximation of the r.h.s. of equations (\ref{nttau}), (\ref{ntau})
should provide a good approximation.

The linear approximation of equations (\ref{nttau}) and (\ref{ntau})
amounts to replace the function $\Psi(y)$ defined in (\ref{psis}) by
the first two terms of the expansion (\ref{psiseries}):
\begin{equation}
\Psi(y)\approx 1-n y~ . \label{oneperone}
\end{equation}
This approximation (\ref{oneperone}) has a transparent probabilistic
interpretation. It corresponds to assuming that any given earthquake
can either trigger no or just one observable aftershock of
first-generation. However, through the avalanche process of
aftershocks which themselves trigger no or just one aftershock, the
total average number $\langle R\rangle = n/(1-n)$ of aftershocks
over all generations which are triggered by a given mainshock can
still be large for $n$ close to $1$. A possible justification of the
approximation that no more than one observable first-generation
aftershock is triggered by a given event is that this statement
refers to observable aftershocks with magnitudes above the threshold
level $m>m_0$. For $m$ large enough above $m_0$, a given mainshock
may trigger many aftershocks of first generation of magnitude larger
than $m_0$, but few of them will be observable with magnitudes
larger than $m$, due to the fast decay of the GR PDF (\ref{GR}) with
increasing $m$. This effect has been shown to lead to a
renormalization of the branching ratio $n$ into an apparent value
$n_a \leq n$ (the equality holding only for the two fixed points
$n=0$ and $n=1$), characterizing the apparent branching structure of
the triggering of observable events (Sornette and Werner, 2005b;
Saichev and Sornette, 2006b).

With the linear approximation (\ref{oneperone}), equation
(\ref{ntau}) takes the form
\begin{equation}
N_-(\tau, m)= Q(m)  +\delta \Phi(\tau) \otimes N_-(\tau,m)~ ,
\label{ntaulin}
\end{equation}
where
\begin{equation}
\delta= n [1-Q^{\gamma-1 \over \gamma}(m)]~ . \label{deltam}
\end{equation}
Similarly, expression (\ref{nttau}) allows us to obtain
\begin{equation}
N(t,\tau, m)=n \Phi(t) \otimes N(t,\tau,m)+ n\Phi(\tau+t) \otimes
N_-(\tau,m)~ . \label{nttaulin}
\end{equation}
Integrating the expression (\ref{oneperone}) term by term over time
yields \be \int_0^\infty N(t,\tau, m)dt= n \int_0^\infty N(t,\tau,
m)dt+ n a(\tau) \otimes N_-(\tau, m)~ , \ee which simplifies into
\begin{equation}
\int_0^\infty N(t,\tau, m)dt= {n \over 1-n} a(\tau) \otimes
N_-(\tau, m)~ . \label{intnttau}
\end{equation}
Using (\ref{intnttau}), we rewrite the probability (\ref{pabs}) in
the form
\begin{equation}
P(\tau)= \exp\left( - {\omega n \over 1-n} a(\tau)\otimes
N_-(\tau,m) - \omega \int_0^\tau N_-(t,m) dt \right)~,
\label{ptransp}
\end{equation}
which now involves only the function $N_-(t,m)$.

Thus, in order to find the explicit expression for $P(\tau)$ from
(\ref{ptransp}), we need to solve equation (\ref{ntaulin}). We can
already state that its solution should obey the following limiting
condition \be \lim_{\tau\to\infty} N_-(\tau,m) = N_-(m)=  {Q(m)
\over 1- \delta}~ . \ee We now introduce the new function
\begin{equation}
g(\tau,m)= {N_-(\tau,m) \over N_-(m)}-1 \quad \iff \quad
N_-(\tau,m)=N_-(m)[1- g(\tau,m)]~ . \label{gnrelation}
\end{equation}
It is easy to show that $g(\tau,m)$ satisfies the equation
\begin{equation}
g(\tau,m)- \delta \Phi(\tau) \otimes g(\tau,m)= \delta a(\tau)~ ,
\label{qeq}
\end{equation}
from which we derive the following relation
\begin{equation}
a(\tau) \otimes g(\tau,m)= A(\tau)+{\delta-1 \over \delta}
\int_0^\tau g(t,m) dt~ . \label{aotimesg}
\end{equation}
Substituting in (\ref{ptransp}) the second relation of
(\ref{gnrelation}) and using (\ref{aotimesg}) yields
\begin{equation}
P(\tau)= \exp\left( -{\omega(m) \tau \over 1- \delta}- {\omega(m)
\over \delta} \Delta\int_0^\tau g(t,m) dt\right)~ , \label{pshort}
\end{equation}
where \be \Delta = {n \over 1-n}- {\delta \over 1-\delta} \ee and
$\omega(m)= \omega Q(m)$ is the average rate of observable
spontaneous events. Recall that $\delta$ is given by (\ref{deltam}).

\begin{figure}[h]
\centerline{\includegraphics[width=14cm]{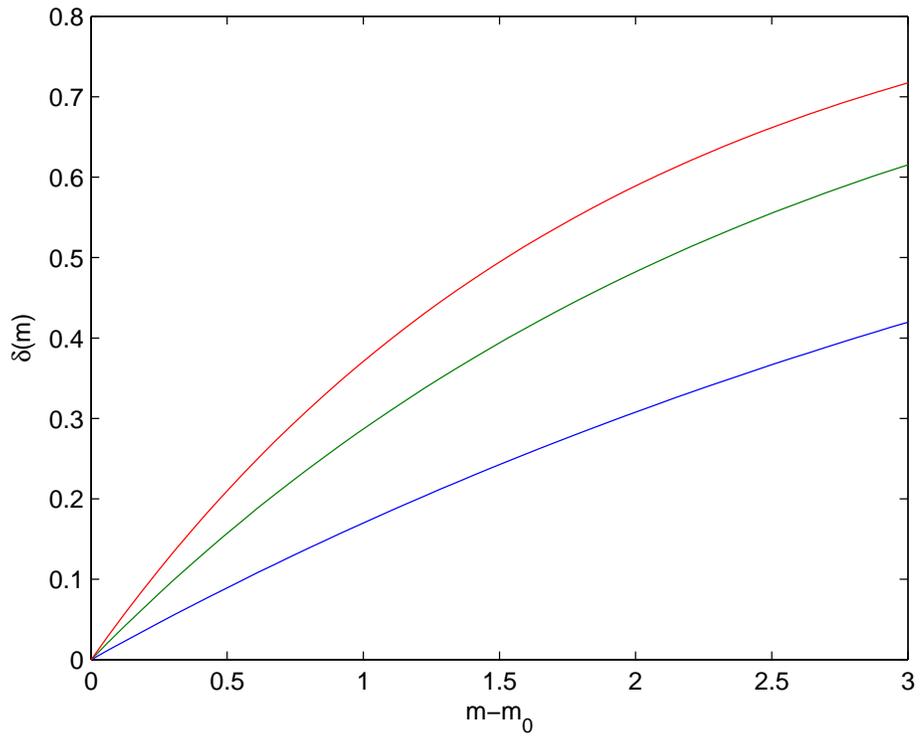}}
\caption{\label{Fig5} Dependence on $m$ of the parameter $\delta$
defined in (\ref{deltam}), which results from the competition
between the GR and productivity laws. Bottom to top:
$\gamma=1.1;1.2;1.3$.}
\end{figure}

A standard technique to obtain the solution of equation (\ref{qeq})
is to use the inverse Laplace transform. It gives
\begin{equation}
g(\tau,m)= g_0\left({\tau \over c},m\right)~ , \label{gexsolution}
\end{equation}
where \be g_0(z,m)=- {\delta\over \pi} \int_0^\infty \text{Im}
\left({ 1- \hat{\Phi}(y,\theta) \over y[1- \delta
\hat{\Phi}(y,\theta)]}\right) e^{-yz} dy \ee and \be
\hat{\Phi}(y,\theta)= y^\theta e^y \Gamma(-\theta,y)~ . \ee Due to
the extremely slow decay of the function $a(\tau)$ (\ref{at}) for
$\theta\ll 1$, the exact solution (\ref{gexsolution}) of equation
(\ref{qeq}) can be described accurately by the quasi-static
approximation
\begin{equation}
g(\tau,m)\approx {a(\tau) \delta \over 1-\delta +\delta a(\tau)}~ .
\label{gsolution}
\end{equation}
Substituting this quasistatic approximation (\ref{gsolution}) in
expression (\ref{pshort}) yields
\begin{equation}
P(\tau)= \exp\left( -{\omega(m) \tau \over 1- \delta}- \omega(m)
\Delta\int_0^\tau {a(t) dt \over 1-\delta +\delta a(t)}\right)~ .
\label{pexplicit}
\end{equation}
It is interesting to notice that, for $\delta\ll 1$, one may neglect
$\delta$ and replace the expression (\ref{pexplicit}) by \be P(\tau)
\approx \exp\left(- \omega(m) \tau- n {\omega(m) \over 1-n}
A(\tau)\right)~ , \label{nbwddw} \ee which coincides with the
simplified model probability (\ref{psimp}) derived in section 3. The
conditions under which this approximation holds can be visualized
from Fig.~\ref{Fig5} showing the dependence of of $\delta$, taking
as a function of $m-m_0$ for different values of $\gamma=b/\alpha$.
This dependence results from the competition between the decay of
the GR law controlled by its exponent $b$ and the value of the
productivity exponent $\alpha$.

Fig.~\ref{Fig5} shows that the condition $\delta \ll 1$ needed to
obtain (\ref{nbwddw}) requires that $m-m_0$ should not be too large,
while Fig.~\ref{Fig4} has shown that the linear approximation
(\ref{oneperone}) used to obtain these results is all the more valid
the larger is $m-m_0$. It is thus worthwhile to come back to the
more complete expression (\ref{pexplicit}) for $P(\tau)$, which
holds for arbitrary values of $\delta$. It is convenient to rewrite
(\ref{averageobs}) so that the average number of windowed events is
equal to
\begin{equation}
x=\lambda(m) \tau~ , \qquad \lambda(m)= {\omega(m) \over 1-n}~ .
\label{averagerate}
\end{equation}
With these notations, we have
\begin{equation}
P(\tau)= \varphi(x,m)= \exp\left(-\eta x- \nu \int_0^x g(y,m,\theta)
dy \right)~ , \label{pconv}
\end{equation}
where
\begin{equation}
g(x,m,\theta)= {\epsilon^\theta \over \delta \epsilon^\theta +
(1-\delta) (\epsilon + y)^\theta}~ . \label{gintegrand}
\end{equation}
and
\begin{equation}
\eta= {1-n \over 1-\delta}~ , \quad \nu= (1-n) \Delta~ , \quad
\epsilon= c \lambda(m)={c \omega(m) \over 1-n}~ . \label{etanu}
\end{equation}
Correspondingly, the sought dimensionless PDF of the recurrence
times defined by \be f(x,m)= {d^2 \varphi(x,m) \over d x^2} \ee is
given by
\begin{equation}
\begin{array}{c}
f(x,m)=\\[3mm]\displaystyle
\left(\theta \nu (1-\delta) \epsilon^{-\theta} (x+
\epsilon)^{\theta-1} g^2(x,m,\theta)+ \left[\eta+ \nu g(x,m,
\theta)\right]^2\right) \varphi(x,m)~ .
\end{array} \label{pdfrectime}
\end{equation}

\begin{figure}[h]
\centerline{\includegraphics[width=14cm]{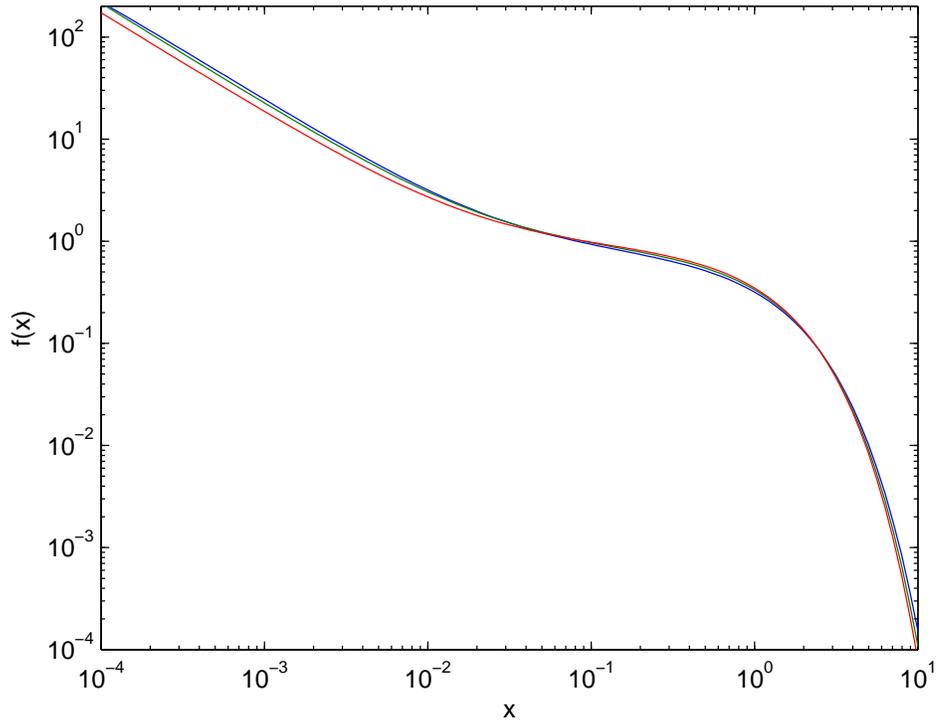}}
\caption{\label{Fig6} Plots of the PDF (\ref{pdfrectime}) of the
scaled recurrence times $x$ for $\theta=0.03$, $\gamma=1.2$, $n=0.9$
and for different magnitude thresholds $m-m_0=2,4,6$.}
\end{figure}

Similarly to the observation that the simplified version
(\ref{pdfrho}) for the PDF of recurrence times depends very slowly
on the magnitude threshold $m$, the PDF of inter-event times
(\ref{pdfrectime}) predicted by the ETAS model also depends weakly
on $m$, if $\theta\ll 1$ and $\gamma$ is close to $1$.
Fig.~\ref{Fig6} plots the dependence of the PDF (\ref{pdfrectime})
as a function of the dimensionless inter-event time $x$ for
$\theta=0.03$, $\gamma=1.2$ and $n=0.9$ for different magnitude
thresholds $m=2, 4, 6$. For such a span of magnitude thresholds, the
average rate (\ref{averagerate}) changes by many orders of
magnitudes, while the corresponding PDF's are amazingly close each
other. Fig.~\ref{Fig6} indeed examplifies the very weak dependence
of the PDF on $m$.

\begin{figure}[h]
\centerline{\includegraphics[width=14cm]{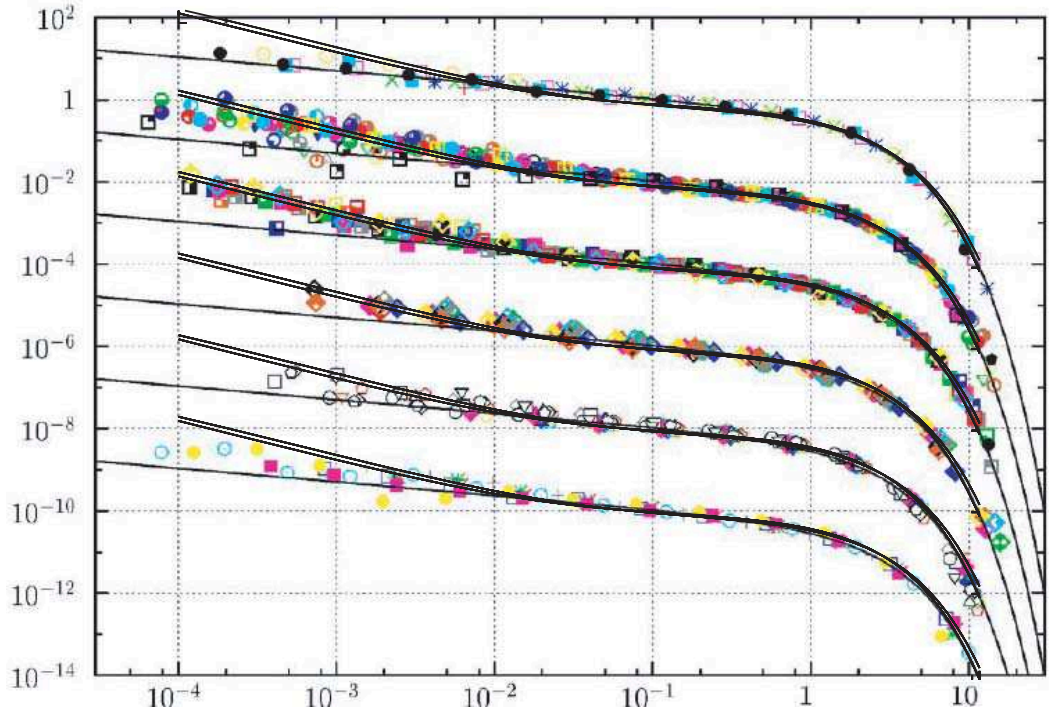}}
\caption{\label{Fig7} Plots of the empirical PDF's of scaled
inter-event times taken from (Corral, 2004a),  Corral's fitting
curves using expression (\ref{Corralfit}) (single solid curves) and
our theoretical prediction (\ref{pdfrectime}) for $\theta=0.03$,
$\gamma=1.2$ and $n=0.9$ (double solid curves). Our ETAS prediction
is shown as double solid curves because we show the theoretical
prediction (\ref{pdfrectime}) for two different magnitude threshold
levels $m-m_0=2$ and $6$, showing the very weak dependence on the
magnitude of completeness. Top to bottom: the NEIC worldwide catalog
for regions with $L \geq 180$ degrees, 1973-2002; NEIC with $L \leq
90$ degrees, (same period of time); Southern California, 1984-2001,
1988-1991, and 1995-1998; Northern California, 1998-2002;
 Japan, 1995-1998, and New Zealand, 1996-2001; (bottom), Spain,
1993-1997, New Madrid, 1975-2002, and Great Britain, 1991-2001. The
curves are translated for clarity by the factors $1; 10^{-2};
10^{-4}; 10^{-6}; 10^{-8}$ and $10^{-10}$ from top to bottom.}
\end{figure}

Fig.~\ref{Fig7} shows the empirical PDF's of the scaled inter-event
times constructed by Corral (2004a), as well as Corral's fitting
curves using expression (\ref{Corralfit}). We superimpose on these
curves our theoretical prediction (\ref{pdfrectime}), which actually
provides a better fit to the data, especially for small $x\lesssim
10^{-2}$, where our ETAS prediction correctly accounts for the
impact of Omori's law, as already mentioned in our discussion of the
simplified model PDF (\ref{pdfrho}). We present the theoretical
prediction (\ref{pdfrectime}) for two different magnitude threshold
levels $m-m_0=2$ and $6$, showing the very weak dependence on the
magnitude of completeness. These results suggest that the parameter
$\theta$ of the (bare) Omori law needs to be small (typically
$\theta \lesssim 0.1$) to account for the data. Similar fits are
obtained for the range of parameters $1 < \gamma \leq 1.2$ and $0.8
\leq n < 1$, in agreement with bounds previously obtained from the
prediction of the ETAS model on the distribution of seismic rates
(Saichev and Sornette, 2006a).

\section{PDF of recurrence times for multiple regions: accounting
for Bak et al. (2002)'s empirical analysis}

Until now, we have derived the prediction of the ETAS model for the
statistics of inter-event times in a single homogeneous region and
have shown that it is compatible with the empirical observation of
an approximate unified scaling law of the form (\ref{unified}). We
have also shown that our theoretical expression fits remarkably well
the empirical PDF's over the whole range of recurrence times,
accounting for different regimes by using only the physics of
triggering quantified by Omori's law.

There is another unified saling law for the statistics of recurrence
time obtained after averaging over multiple regions, which has been
proposed by Kossobokov and Mazhkenov (1988) and Bak et al. (2002).
We will refer for short to this law as Bak et al.'s unified law. The
aim of this section is to derive analytically Bak et al.'s unified
law obtained for multiple regions based on the prediction of the
ETAS model that we have presented in the previous section and on the
simple hypothesis that all regions obey the PDF form
(\ref{pdfrectime}) but with different average seismic rates. This
simply means that Bak et al.'s unified law in our view just results
from an appropriate averaging of expression (\ref{pdfrectime}) over
the statistics of the average rates of the multiple regions under
observation.

We first present in subsection \ref{grasda} a general scheme to
perform this averaging in the next subsection with various test
statistics for the average rates in different regions. Then, in the
second subsection \ref{grasda2}, we introduce a fractal geometrical
model which allows us to propose a specific statistics for the
average rates in different regions, which lead to excellent fits to
the empirical data.

\subsection{General statistics for the average regional seismic rates \label{grasda}}

In order to choose reasonable statistics for the average seismic
rates $\omega_i$ in multiple regions with the minimum of additional
parametes, we assume that these $\omega_i$'s obey a principle of
statistical self-similarity: if the branching ratios $n_i$'s and the
linear size $L_i$'s of different regions are identical, then the
seismic rates can be written
\begin{equation}
\omega_i= \bar{\omega} ~u_i~ , \label{omegavar}
\end{equation}
where $\bar{\omega}$ is the rate of spontaneous events averaged over
all regions, while the $u_i$'s are mutually independent random
variables distributed according to the same PDF $\mathcal{E}(u)$.
The PDF $\mathcal{E}(u)$ is assumed to be independent of the common
values of the branching ratio $n$ and linear size $L$ of the
multiple regions. Additionally, we assume that, as for the scaling
law (\ref{unified}) of a single region, the PDF $\mathcal{E}(u)$
does not depend on the magnitude threshold $m$ (which is common to
all regions). One may interpret this assumption as a consequence of
the scaling properties of the GR law.

Since the average rates of spontaneous events cannot be directly
observed, it may be preferable to replace the statement
(\ref{omegavar}) by the equivalent representation
\begin{equation}
\lambda_i= \bar{\lambda}~ u_i~ , \label{lambdavar}
\end{equation}
which involves the total observable seismic rates above a magnitude
threshold. Due to (\ref{lambdavar}), the PDF $\mathcal{E}(u)$ should
satisfy to double normalization condition
\begin{equation}
\int_0^\infty \mathcal{E}(u)du = \int_0^\infty u \mathcal{E}(u) du
\equiv 1 ~ . \label{doublenorm}
\end{equation}
In this subsection, we concentrate our attention on the universal
properties of the PDF of recurrence times constructed over multiple
regions, which are found almost the same for qualitatively different
distributions $\mathcal{E}(u)$. In the next subsection
\ref{grasda2}, we will obtain a specific expression for the PDF
$\mathcal{E}(u)$ from a simple model based on a fractal distribution
of spontaneous earthquake sources.

Applying our conjecture (\ref{lambdavar}) to the general relation
(\ref{d2pdf}), the PDF of inter-event times over multiple regions is
obtained as
\begin{equation}
\mathcal{H}(\tau)= {1 \over\rule{0mm}{4mm} \bar{\lambda}} {d^2 \over
d\tau^2} \int_0^\infty P(\tau) \mathcal{E}(u) {du \over u}~ , \qquad
\bar{\lambda}= {\bar{\omega} \over 1-n}~ . \label{multiPDF}
\end{equation}
It is convenient to represent the probability $P(\tau)$ given by
(\ref{pabs}) that no events occur in a given single region in the
form
\begin{equation}
P(\tau)= e^{-u K(x)}~ , \label{singlePDF}
\end{equation}
where the auxiliary function \be K(x)= \bar{\omega} L\left({x \over
\bar{\lambda}},m\right)~ , \qquad x= \bar{\lambda} \tau~ , \ee does
not depend on $u$. It follows from (\ref{pexplicit}) that \be K(x)=
\eta x+ \nu \int_0^x g(y,m,\theta) dy ~ , \ee where $g(y,m,\theta)$
is given by (\ref{gintegrand}). Substituting (\ref{singlePDF}) into
(\ref{multiPDF}) yields the following expression analogous to
(\ref{1}) for the PDF of the recurrence times over multiple regions:
\begin{equation}
\mathcal{H}(\tau)= \bar{\lambda} h(\bar{\lambda} \tau)
\label{multiscale}
\end{equation}
where
\begin{equation}
h(x)= \left({d K(x)\over dx}\right)^2 \hat{\mathcal{E}}_1(K(x))-
{d^2 K(x)\over dx^2} \hat{\mathcal{E}}(K(x)) ~ . \label{hdimless}
\end{equation}
We have used the following Laplace transforms
\begin{equation}
\hat{\mathcal{E}}(x)= \int_0^\infty \mathcal{E}(u) e^{-ux} du~ ,
\quad \hat{\mathcal{E}}_1(x)= - {d \hat{\mathcal{E}}(x) \over dx}=
\int_0^\infty u \mathcal{E}(u) e^{-ux} du~ . \label{elapim}
\end{equation}

It is possible to derive the following asymptotic behavior of the
PDF $h(x)$ (\ref{hdimless}). Below, we will illustrate the main
features of these asymptotics using the limiting case $\delta=0$ and
$\tau\gg c$ such that $K(x)$ reduces to
\begin{equation}
K(x)\approx (1-n) x+ {n \over 1-\theta} \epsilon^\theta
x^{1-\theta}~ ,\label{kx}
\end{equation}
corresponding to the simplified model of recurrence time statistics.

Let us first consider the behavior of $h(x)$ for $x\ll 1$. First,
the asymptotic behavior of the dimensionless PDF $h(x)$ for $x\ll 1$
is universal in the sense that it does not depend on the shape of
the PDF $\mathcal{E}(u)$ of the average rates of the multiple
regions. Indeed, due to the double normalization condition
(\ref{doublenorm}), we have \be \mathcal{E}(K)\simeq
\mathcal{E}_1(K) \simeq 1 \qquad (K\ll 1)~. \ee Thus, in view of
(\ref{hdimless}) and (\ref{kx}), we obtain
\begin{equation}
h(x) \approx (1-n+ n \bar{\epsilon}^{\;\theta} x^{-\theta})^2 +
\theta n \bar{\epsilon}^{\;\theta} x^{-1-\theta} \qquad (x\ll 1)~ .
\label{hxsmall}
\end{equation}
For very small $x$'s, the last term prevails and we have the
universal asymptotic
\begin{equation}
h(x) \approx \theta n \bar{\epsilon}^{\;\theta} x^{-1-\theta}~ ,
\qquad x\ll x_* \ll 1~ . \label{hvsmall}
\end{equation}
The range of validity of this asymptotic can be roughly estimated by
putting formally $\bar{\epsilon}^{\;\theta} x^{-\theta}\approx 1$ in
the r.h.s. of relation (\ref{hxsmall}), which gives
\begin{equation}
h(x)\approx 1+ \theta n x^{-1} \label{hxvsm}
\end{equation}
so that \be x_* \approx \theta n~ . \ee

Let us now consider the behavior of $h(x)$ for $x\gg 1$. We observe
an almost universal asymptotic of the dimensionless PDF $h(x)$ given
by (\ref{hdimless}) for $x\gg 1$. Indeed, let us assume that the PDF
of the average seismic rates of the multiple regions has, for small
$u$, the following power asymptotic
\begin{equation}
\mathcal{E}(u) \approx \mu u^\alpha~ , \qquad u\ll 1~ .
\label{smallu}
\end{equation}
This leads to the explicit form of the Laplace transforms
(\ref{elapim}): \be \hat{\mathcal{E}}(x) \simeq \mu \Gamma(1+\alpha)
x^{-1-\alpha}~ , \qquad \hat{\mathcal{E}}_1(x) \simeq \mu
\Gamma(2+\alpha) x^{-2-\alpha}~ , \qquad x\gg 1~ . \ee This finally
yields \be
\begin{array}{c}
h(x) \approx \mu \Gamma(1+\alpha) \times \\[3mm]
\left[(1+\alpha)(1-n +n \bar{\epsilon}^{\;\theta} x^{-\theta})^2
K^{-2-\alpha}(x) +\theta n \bar{\epsilon}^{\;\theta} x^{-1-\theta}
K^{-1-\alpha}(x) \right]~ .
\end{array}
\ee For small $\theta$, the leading behavior of this asymptotic can
be obtained by using $K(x) \approx x$, which gives
\begin{equation}
h(x) \sim x^{-2-\alpha}~ . \label{hxgr}
\end{equation}

\begin{figure}[h]
\centerline{\includegraphics[width=14cm]{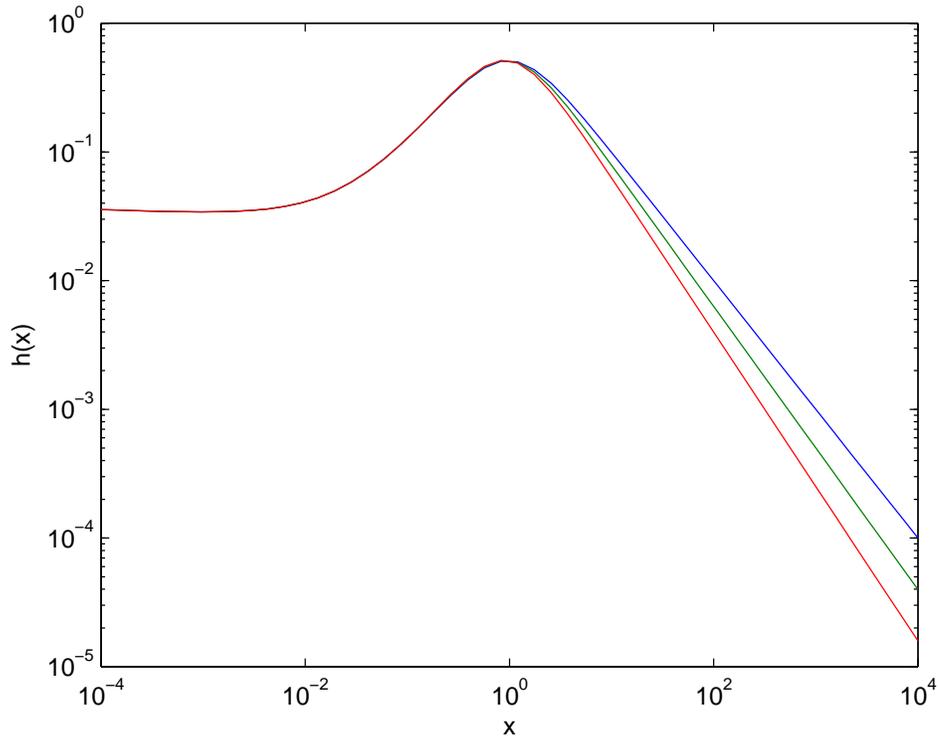}}
\caption{\label{Fig8} Plots of the function $x h_m(x)$ for
$\theta=0.03$, $n=0.9$ and for different values of
$\alpha=0;0.1,0.2$, demonstrating the approximate unified scaling
law of the PDF of inter-event times obtained over multiple regions,
as reported by Bak et al. (2002).}
\end{figure}

A useful overview of the behavior of the dimensionless PDF $h(x)$
valid for all interdiate values $x$, including the two above
asymptotics (\ref{hvsmall}) and (\ref{hxgr}), can be obtained by
considering a smoothed function interpolating between them. Let us
for instance consider the following modeling function
\begin{equation}
h_m(x)= {1+ \theta n x^{-1-\theta} \over 1+ x^{2+\alpha}}~ .
\label{hmod}
\end{equation}
We put here for simplicity $\mu \Gamma(2+\alpha)\approx 1$.
Fig.~\ref{Fig8} shows the function $x h_m(x)$ for $\theta=0.03$,
$n=0.9$ and for different values $\alpha$, which mimics the
empirical construction reported by Bak et al. (2002). Note in
particular the existence of a plateau for $x\ll 1$, corresponding to
the Omori law asymptotic (\ref{hvsmall}). One can also observe the
power law asymptotic $x^{-1-\alpha}$, corresponding to the
asymptotic (\ref{hxgr}) for $x\gg 1$. In addition, there is an
intermediate kink for $x\simeq 1$, due to the first term of the
asymptotic formula (\ref{hxvsm}). All three features mimic the
behavior found by Bak et al. (2002).

\begin{figure}[h]
\centerline{\includegraphics[width=14cm]{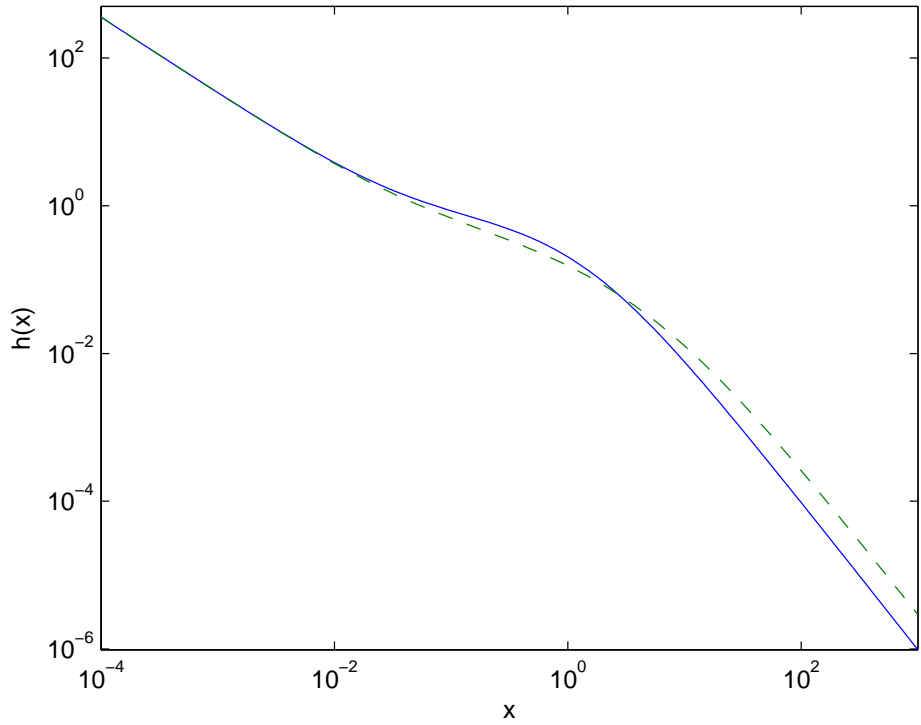}}
\caption{\label{Fig9} Plots of the PDF $h(x)$ given
by(\ref{hdimless}) for $\theta=0.03$, $n=0.9$, obtained by using the
two examples (\ref{gammadist}) and (\ref{powertail}) for the the PDF
of the average seismic rates of the multiple regions. Solid line:
the Gamma distribution (\ref{gammadist}) for $\alpha=0$; dashed
line: the power tail distribution (\ref{powertail})  for $p=0.5$.}
\end{figure}

These properties of $h(x)$  are weakly dependent on the value of the
exponent $\alpha$, which controls the asymptotic behavior
(\ref{smallu}) of the PDF of the average seismic rates of the
multiple regions for small rates. To illustrate this point, consider
two qualitatively different sample distributions $\mathcal{E}(u)$
which obey the double normalization (\ref{doublenorm}). Our first
example is the Gamma distribution formulated so that it obeys
(\ref{doublenorm}):
\begin{equation}
\mathcal{E}^g(u)= {(1+\alpha)^{1+\alpha} \over \Gamma(1+\alpha)}
u^\alpha e^{-(1+\alpha) u}~ . \label{gammadist}
\end{equation}
The main properties of this distribution are the presence of a power
asymptotic (\ref{smallu}) for small $u$'s and an exponential decay
for $u\gg 1$. In this case, the Laplace transforms (\ref{elapim})
are \be \hat{\mathcal{E}}^g(x)= \left({1+ \alpha \over 1+\alpha +x}
\right)^{1+\alpha}~ , \qquad \hat{\mathcal{E}}_1^g(x)= \left({1+
\alpha \over 1+\alpha +x} \right)^{2+\alpha}~ . \ee Our second
example for a distribution obeying (\ref{doublenorm}) is
\begin{equation}
\mathcal{E}^p(u)= {p+1 \over p} \left(1+ {u \over p}\right)^{-2-p}~,
\label{powertail}
\end{equation}
which corresponds to the particular case $\alpha=0$. The main
difference between this distribution and the previous Gamma
distribution is the slowly decaying power law tail for $u\gg 1$: \be
\mathcal{E}^p(u) \sim u^{-p-2} \qquad (u\gg 1)~ . \ee In this case,
the Laplace transforms (\ref{elapim}) read \be
\hat{\mathcal{E}}^p(x)= (1+p) (px)^{1+p} e^{px} \Gamma(-1-p, px) \ee
and \be \hat{\mathcal{E}}_1^p(x)= {1\over x} \left(1+p-(1+p+ px)
\mathcal{E}_p(x)\right)~ . \ee Fig.~\ref{Fig9} plots the PDF $h(x)$
obtained with the two examples (\ref{gammadist}) and
(\ref{powertail}). It would be difficult to distinguish between
these two distributions in a realistic empirical setting with
limited data and large statistical fluctuations.

\begin{figure}[h]
\centerline{\includegraphics[width=14cm]{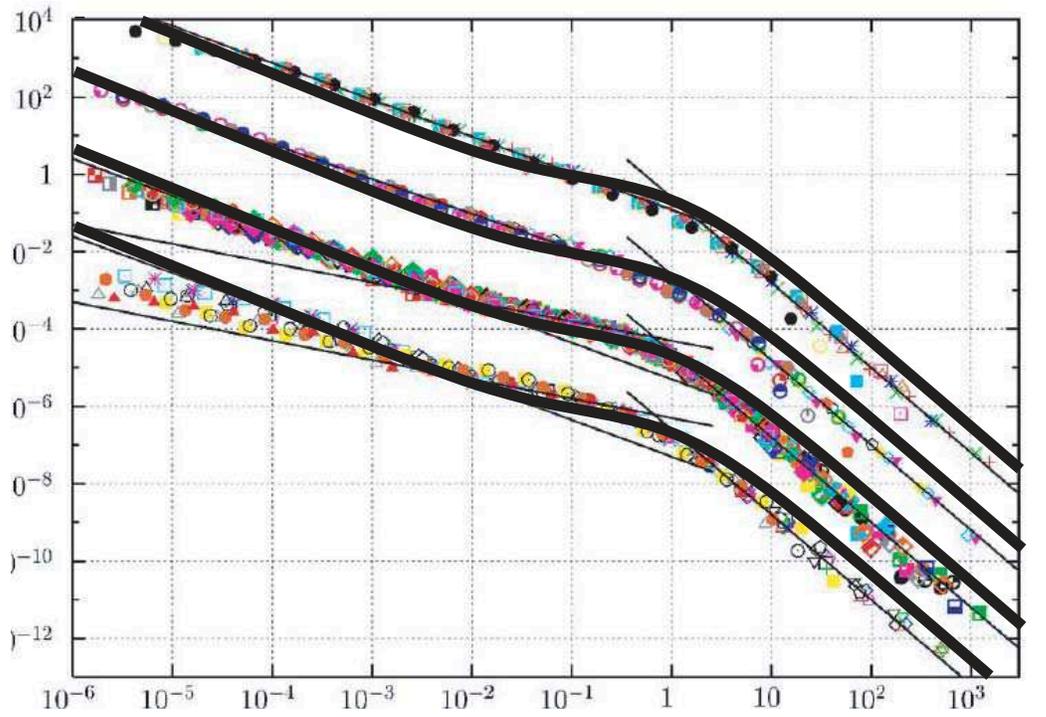}}
\caption{\label{Fig10} Empirical PDF's of the recurrence times
between earthquakes over multiple regions following Bak et al.
(2002)'s procedure obtained from Fig.~2 of (Corral, 2004a), on which
has been superimposed our prediction for $h(x)$ obtained with the
Gamma distribution (\ref{gammadist}) for $\mathcal{E}(u)$ with
$\theta=0.03$, $n=0.9$, $\epsilon=0.76$ and $\alpha=0.2$. The curves
have been translated from top to bottom by the factors
$1;10^{-2};10^{-4};10^{-6}$. }
\end{figure}

In the next subsection \ref{grasda2}, we develop a simple model for
$\mathcal{E}(u)$ based on the idea that seismic rates of spontaneous
events are distributed on a fractal geometrical set. In a nutshell,
the key idea is the following: a spatial distribution of spontaneous
events with fractal dimension $d<2$ implies that, the larger is the
number of events, the larger are the spatial distances between them.
Thus, the PDF of the number $R$ of events within a given region
should decay quite fast at the number of events increases. This
tends to favor the decay for large seismic rates modelled by the
Gamma distribution (\ref{gammadist}) over the power law distribution
(\ref{powertail}). To finish this general discussion, we present in
Fig.~\ref{Fig10} the empirical data analyses by Corral (2004a)
together with prediction for $h(x)$ obtained by using the Gamma
distribution for $\mathcal{E}(u)$ with $\theta=0.03$, $n=0.9$,
$\epsilon=0.76$ and $\alpha=0.2$. The fits to the empirical data are
suggestive as they captures all its qualitative features. There are
however undeniable quantitative differences which result in part
from the fact that the Gamma distribution is probably not the true
distribution for $\mathcal{E}(u)$ even if it captures correctly the
mean properties of the data.

\subsection{Probabilistic consequences of the fractal geometry
of the spatial distribution of spontaneous earthquakes
\label{grasda2}}

In the framework of the ETAS model combined with the hypothesis of
statistical self-similarity, the preceding subsection \ref{grasda}
has stressed that an analytical understanding of Bak et al.'s
unified law requires the knowledge of the PDF $\mathcal{E}(u)$ of
the average rates over the multiple regions. To constrain its
structure, we make the assumption that the well-known fractal
spatial organization of all observed earthquakes reflects a similar
fractal spatial organization of the subset of spontaneous events
which are the origin of the earthquake triggering activity. We thus
assume that spontaneous events are distributed spatially on a
fractal geometry, which we are going to exploit to derive a natural
form for the PDF $\mathcal{E}(u)$.

For this goal, the key idea is based on an important relation
between two reciprocal random variables. The first random variable
is the area $S(k)$ occupied by $k$ spontaneous events which occurred
during a fixed time interval of duration $\tau$. The second random
variable $k(S)$ is the inverse function of $S(k)$, which is nothing
but the number of spontaneous events within the given area $S$.
Although both functions are ill-defined in practice due to the
difficulties in defining unambiguously the areas occupied by
chaotically occurring disseminated events, we can nevertheless
obtain approximate scaling laws for the distributions of these two
variables, which derive from the fact that the average $k(S)$, which
is proportional to the rate of spontaneous events $\omega(s)$,
should obey to power law
\begin{equation}
\langle k(S)\rangle \sim \omega(S) \sim L^d~ . \label{fracscale}
\end{equation}
Here, $L$ is a characteristic scale of the area $S$ and $d$ is the
fractal dimension of the spatial set of spontaneous events.

In order to obtain these scaling laws, we recognize that the $i$-th
spontaneous event ``occupies'' a spatial area $s_i$ such that, if
some area $S$ contains $k$ spontaneous events, then it is equal to
\begin{equation}
S(k)\approx s_1+ s_2+ \dots + s_k~ . \label{sumk}
\end{equation}
One may interpret $s_i$ as given by $s_i\approx L_i^2$ where $L_i$
is the distance between the $i$-th event and its nearest neighbor.
Next, let us assume that the summands in (\ref{sumk}) are
statistically independent with a common PDF $v(s)$ which has the
following aymptotic power law \be v(s) \approx \eta {d \over
\Gamma(1-d)} s^{-1-g}~, \qquad s\to \infty \qquad  0<g<1~ .
\label{mgmgle} \ee The expression (\ref{mgmgle}) corresponds to a
distribution of spatial jumps, as in a so-called L\'evy flights
(Metzler and Klafter, 2000), which underlies the fractal spatial
distribution of spontaneous events. From the generalized central
limit theorem (Gnedenko and Kolmogorov, 1954; Sornette, 2004), the
PDF of the random sum (\ref{sumk}) for $k\gg 1$ is given
asymptotically by
\begin{equation}
V_k(s)\approx {1 \over (\eta k)^{1/g}} f_g \left( {s \over (\eta
k)^{1/g}}\right)~ , \qquad k\gg 1~ , \label{infdiv}
\end{equation}
where $f_g(x)$ is an infinitely divisible distribution, whose
Laplace transform is equal to \be \hat{f}_g(y)= \int_0^\infty f_g(x)
e^{-yx} dx= e^{-v^g}~ , \qquad 0<g<1~ . \ee

Let us now view expression (\ref{sumk}) from a different vantage.
One can also interpret (\ref{sumk}) as an implicit equation \be
S(k)=S \ee for the unknown variable equal to the number $k(S)$ of
spontaneous events occurring within the given area $S$. Piryatinska
et al. (2005) have shown that the PDF of the random variable $S$
allows one to derive the PDF of the random variable $k(S)$ as
\begin{equation}
W(k;S)= {\eta \over S^g} w_g \left({\eta k \over S^g} \right)~ ,
\label{invinfdiv}
\end{equation}
where
\begin{equation}
w_g(x)= {1 \over g x^{1+1/g}} f_g \left({1 \over x^{1/g}} \right)~ .
\end{equation}
We then deduce that the average of the random variable $k(S)$
distributed according to the PDF (\ref{invinfdiv}) is given by \be
\langle k(S)\rangle = {S^g \over \eta \Gamma(1+g)} \label{mbndw} \ee

\begin{figure}[h]
\centerline{\includegraphics[width=14cm]{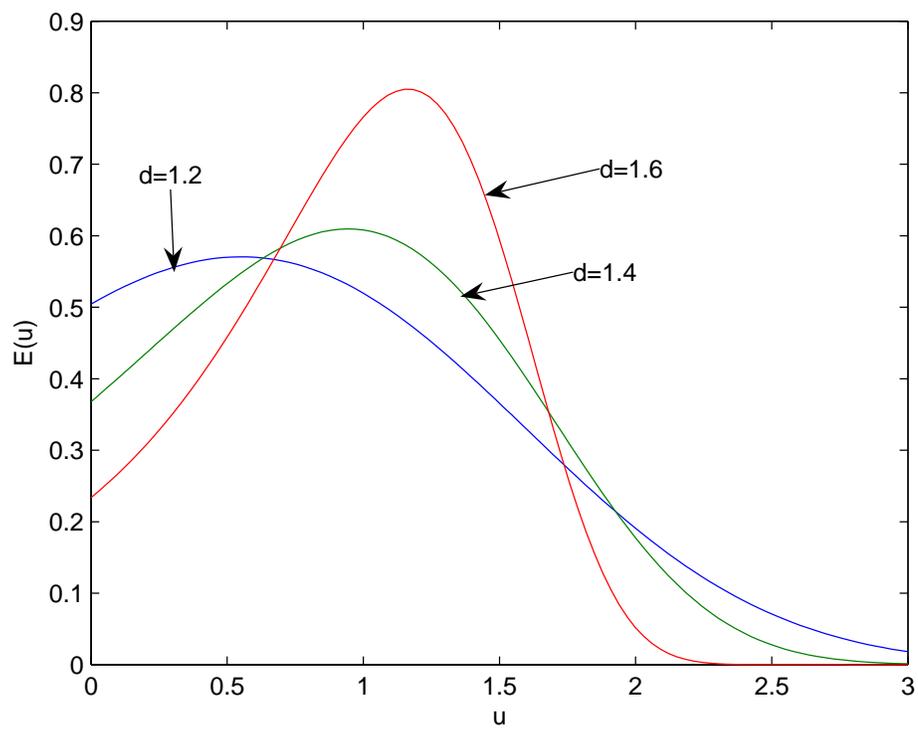}}
\caption{\label{Fig11} Plots of the distribution $\mathcal{E}(u)$
given by (\ref{fracpdf}) for $d=1.2;1.4;1.6$.}
\end{figure}

Since $S\approx L^2$ in (\ref{mbndw}) and comparing with
(\ref{fracscale}) yields the self-consistent determination of the
exponent $g$ \be g=d/2~, \ee which is indeed between $0$ and $1$ for
$0 < d < 2$. Assuming that average rates of single regions are
proportional to $k(S)$, that is $\lambda \sim \omega\sim k(S)$, we
obtain \be \lambda_i =\bar{\lambda} u_i~ , \qquad u_i= {k(S_i) \over
\langle k(S) \rangle}~ , \ee where $S_i$ are different regions with
the same linear size $L$. The sought PDF of the average seismic
rates is thus equal to
\begin{equation}
\mathcal{E}(u)= {1 \over \bar{u}} w_{d/2} \left( {u \over
\bar{u}}\right)~ , \qquad \bar{u}= \Gamma(1+d/2)~ . \label{efracpdf}
\end{equation}
One can show that
\begin{equation}
\begin{array}{c}
w_g(u)= \\[3mm]
\displaystyle {1 \over \pi g} \int_0^\infty \exp\left[ x^{1/g}
\cos\left({\pi \over 2g}\right)\right] \cos\left[ u x- x^{1/g}
\sin\left({\pi \over 2g}\right)\right]dx~ , \qquad 0.5<g<1~ .
\end{array} \label{fracpdf}
\end{equation}
Fig.~\ref{Fig11} plots the PDF $\mathcal{E}(u)$ given by
(\ref{fracpdf}) for different values of the fractal dimension $d$ of
the set of earthquake epicenters. The tail of $\mathcal{E}(u)$ is
more extended for smaller fractal dimensions $d$.

\begin{figure}[h]
\centerline{\includegraphics[width=11cm]{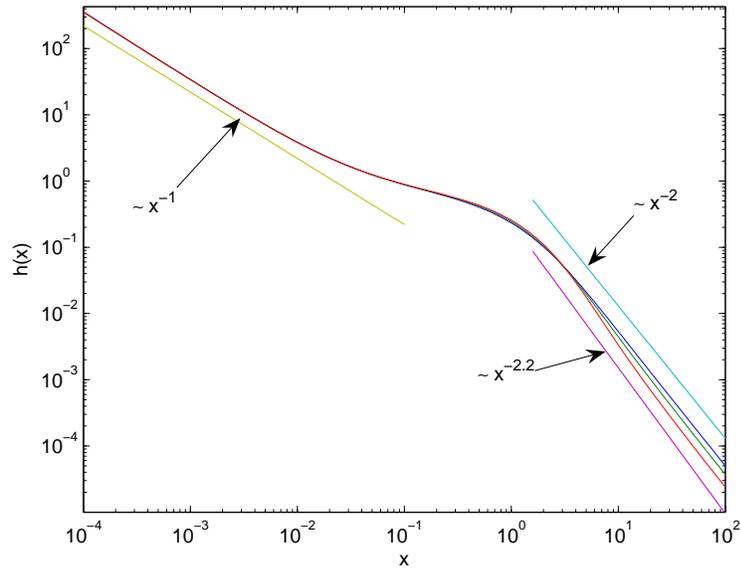}}
\caption{\label{Fig12} Plots of the PDF $h(x)$ of inter-event times
given by (\ref{hdimless}) corresponding to the PDF (\ref{efracpdf})
of the average regional seismic rates. Top to bottom on the right
side of the picture $d=1.2;1.4;1.6$. The straight lines show
asymptotic power laws, as proposed by Bak et al. (2002). }
\end{figure}

The explicit form (\ref{efracpdf}) of $\mathcal{E}(u)$ allows us to
obtain the Laplace transforms $\hat{\mathcal{E}}(x)$ defined by
(\ref{elapim}) as \be \hat{\mathcal{E}}(x)= E_{d/2}(-\bar{u} x)~ ,
\ee where $E_g(z)$ is the Mittag-Leffler function which has the
following integral representation \be E_g(-v)= {v \over \pi}
\sin(\pi g) \int_0^\infty {y^{g-1} e^{-y} dy \over v^2+y^{2g}+ 2 v
y^g \cos(\pi g)}~ . \ee We then calculate explicitely the
corresponding PDF $h(x)$ given by (\ref{hdimless}) of the
inter-event times associated with (\ref{efracpdf}). The results are
shown in Fig.~\ref{Fig12} for three different values $d=1.2, 1.4,
1.6$ of the fractal dimension of earthquake epicenters.

\begin{figure}[h]
\centerline{\includegraphics[width=11cm]{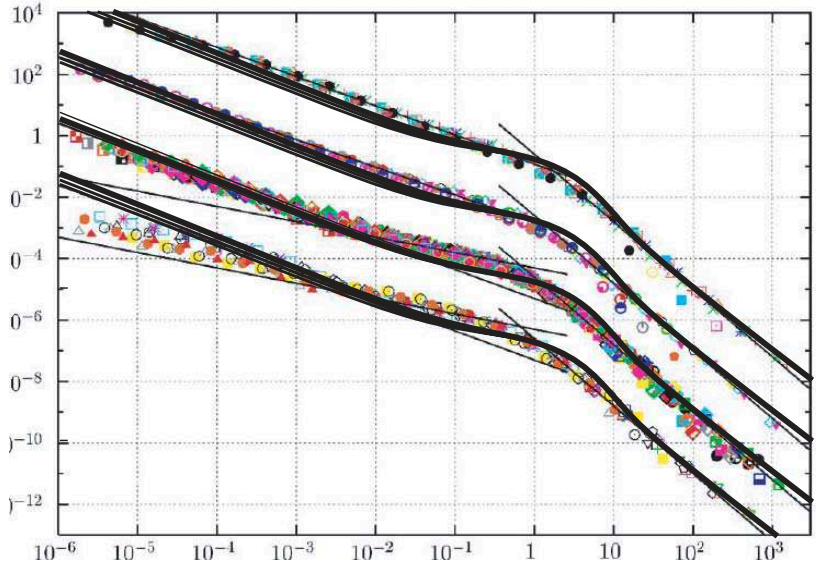}}
\caption{\label{Fig13} Plots of the empirical PDF of inter-event
times in multiple regions collected in Fig.~2 of (Corral, 2004a),
which is compared with our theoretical PDF $h(x)$ given by
(\ref{hdimless}) (thick lines), represented for the parameters
$d=1.8$, $n=0.9$, $\gamma=1,1$ and three values of the threshold
magnitude $m-m_0=2;4;6$. Corral calculated the recurrence-time
probability densities following Bak et al. (2002)'s procedure, after
rescaling by the regional seismic rates. 84 distributions are shown,
with region sizes $L$ ranging from $0.039^{\circ}$ to $45^{\circ}$,
and threshold magnitudes $m$ varying between $1.5$ and $6$. The
curves are shifted by factors $1; 10^{-2}; 10^{-4}$ and $10^{-6}$,
and correspond to: 1 (top), Southern California, 1984-2001; 2,
Northern California, 1985-2003; 3, Southern California, 1988-1991
(stationary rate), NEIC, 1973-2002, Japan, 1995-1998, and Spain,
1993-1997; 4 (bottom), New Zealand, 1996-2001, and New Madrid,
1975-2002. The thin straight lines are Corral's fits by power laws
of the short or intermediate times: 1 (top), $0.12/x^{0.95}$; 2,
$0.15/x^{0.9}$, 3 and 4 (bottom), $0.05/x^{0.95}$ and $0.5/x^{0.5}$.
In all cases, Corral has fitted the long-time tail by
$0.25/x^{2.2}$. The times in the horizontal axis span from 2 min to
about 20 years. Recurrence times smaller than 4, 10, and 2 min are
not shown, for Japan, NEIC, and the rest of catalogs, respectively.}
\end{figure}

Fig.~\ref{Fig13} is the culmination of the present work. It first
makes use of our theoretical analysis based on the ETAS model
leading to the prediction (\ref{pdfrectime}) for a single
homogeneous seismc region, which is combined for multiple regions
with different seismic rates distributed according to the PDF
(\ref{efracpdf}) predicted from our fractal model of earthquake
epicenters, to finally obtain the global PDF (\ref{hdimless}) of
inter-event times. Our final prediction (\ref{hdimless}) is compared
to the empirical distribution represented in Fig.~2 of (Corral,
2004a), for the parameters $d=1.8$, $n=0.9$, $\gamma=1,1$ and to 3
values of the magnitude threshold $m-m_0=2;4;6$. The agreement is
remarkable.

\section{Concluding remarks}

We have proposed a general theoretical approach to describe the
statistics of recurrence times between successive earthquakes in a
given region, or averaged over multiple regions. This work was
motivated by the reports by several authors that recurrence times
between successive earthquakes should be considered for broad areas,
rather than for individual faults, and could provide important
insights in the physical mechanisms of earthquakes. Several authors
mainly from the Physics literature have proposed that the scaling
law (\ref{1}) found to describe well empirical data reveals a
complex spatio-temporal organization of seismicity, which can be
viewed as an intermittent flow of energy released within a
self-organized (perhaps critical) system, for which concepts and
tools from the theory of critical phenomena can be applied.

We have shown that this view is probably too romantic because much
simpler explanations can be proposed to fully account for the
empirical observations. Indeed, we have shown that the so-called
universal scaling laws of inter-event times do not reveal more
information than what is already captured by the well-known laws of
seismicity (Gutenberg-Richter and Omori, essentially), together with
the assumption that all earthquakes are similar (no distinction
between foreshocks, mainshocks and aftershocks). This conclusion is
reached by a combination of analyses, which start from a
generalization of Molchan (2005)'s argument, to go to simple models
of triggered seismicity taking into account Omori's law, and end
with a detailled study of what the ETAS model has to tell us on the
statistics of inter-event times. By using the formalism of
generating probability functions, we have been able to derive
analytically specific predictions for the PDF of recurrence times
between earthquakes in a single homogeneous region as well as for
multiple regions. Our theory has been found to account
quantitatively precisely for the empirical power laws found by Bak
et al. (2002) and Corral (2003; 2004a). We showed in particular that the empirical
statistics of inter-event times result from subtle cross-overs
rather than being genuine asymptotic scaling laws. We also showed
that universality does not strictly hold.

Therefore, to the question raised in the introduction, we are led to
conclude that the statistics on inter-event times described in (Bak
et al., 2002; Corral, 2003, 2004a,b, 2005a; Livina et al., 2005) is
not really new in the sense that they do not reveal information which
is not already contained in the known laws of seismicity. Our
conclusion is that they can be derived from the known
statistical properties of seismicity, so that they are only
different ways of presenting the same information. In particular,
the fact that the simple models of Lindman et al. (2005) are not
able to fully reproduce the structure of the empirical statistics of
recurrence times, as noted by Corral and Christensen (2006), does
not necessarily imply that the so-called ``unified scaling laws''
reveal any novel information. We think we have clearly shown that
they can be derived from simple and well-known physical ingredients,
when carefully taking into account the physics of triggering between
earthquakes. In this sense, the present work is the continuation of
an effort to classify the empirical observations which can be, from
those which cannot be, explained by the simple ETAS benchmark
(Helmstetter and Sornette, 2003a; 2003b; Saichev and Sornette, 2005;
2006a). With this effort, we hope to eventually help identify real robust
statistics which can not be explained by the ETAS benchmark or its
siblings, leading us towards the acquisation of new interesting and
important knowledge on the physics of earthquakes with potential
applications for their forecasts.

\pagebreak

{\bf References} \vskip 1cm

Bak, P., K. Christensen, L. Danon, and T. Scanlon (2002), Unified
Scaling Law for Earthquakes, Phys. Rev. Lett. 88, 178501.

Christensen, K., L. Danon, T. Scanlon, and P. Bak (2002), Unified
Scaling Law for Earthquakes Proc. Natl. Acad. Sci. USA 99, 2509-2513
(2002).

Console, R., A.M. Lombardi, M. Murru, D. Rhoades (2003a), Bath's law
and the self-similarity of earthquakes, J. Geoph. Res., 108,
2128-2136, doi:10.1029/2001JB001651.

Console, R. and Murru, M., (2001), A simple and testable model for
earthquake clustering, J. Geophys. Res. 106, 8699-8711.

Console R., Murru, M., Lombardi, A.M., (2003b), Refining earthquake
clustering models, J. Geophys. Res. 108, 2468,
doi:10.1029/2002JB002130.

Console, R., M Murru, and F. Catalli (2006), Physical and stochastic
models of earthquake clustering, Tectonophysics 417, 141-153

Console, R., D. Pantosti, G. D'Addezio (2002), Probabilistic
approach to earthquake prediction, Ann. Geophysics, 45 (6), 723-731.

Corral, A. (2003), Local distributions and rate fluctuations in a
unified scaling law for earthquakes, Phys. Rev. E 68, 035102(R).

Corral, A. (2004a), Universal local versus unified global scaling
laws in the statistics of seismicity, Physica A, 340, 590-597.

Corral, A. (2004b), Long-term clustering, scaling, and universality
in the temporal occurrence of earthquakes, Phys. Rev. Lett. 92,
108501.

Corral, A. (2005a), Mixing of rescaled data and Bayesian inference
for earthquake recurrence times, Nonlinear Processes in Geophysics
12, 89-100 (2005).

Corral, A. (2005b), Renormalization-Group Transformations and
Correlations of Seismicity,
 Phys. Rev. Lett. 95, 028501.

Corral, A. and K. Christensen (2006), Comment on ``Earthquakes
Descaled: On Waiting Time Distributions and Scaling Laws,'' Phys.
Rev. Lett. 96, 109801.

Daley, D.J. and D. Vere-Jones (1988), An Introduction to the Theory
of Point Processes, New York, Berlin: Springer-Verlag.

Davy, P., A. Sornette and D. Sornette (1990), Some consequences of a
proposed fractal nature of continental faulting, Nature 348, 56-58.

Gerstenberger, M.C., S. Wiemer, L.M. Jones and P.A. Reasenberg,
(2005), Real-time forecasts of tomorrow's earthquakes in California,
Nature 435 (7040), 328-331.

Global Earthquake Satellite System -- A 20 year plan to enable
earthquake prediction (2003), NASA and JPL, JPL 400-1069 03/03.

Gnedenko, B.V. and A. N. Kolmogorov (1954), Limit Distributions for
Sums of Independent Random Variables (Addison-Wesley).

Hawkes, A.G. and D. Oakes (1974), A cluster representation of a
self-excited point process, J. Appl. Prob. 11, 493-503.

Helmstetter, A., Y. Kagan and D. Jackson (2005), Importance  of
small earthquakes for stress transfers and earthquake triggering, J.
Geophys. Res., 110, B05S08, 10.1029/2004JB003286.

Helmstetter, A. and D. Sornette (2002), Sub-critical and
supercritical regimes in epidemic models of earthquake aftershocks,
J. Geophys. Res. 107, NO. B10, 2237, doi:10.1029/2001JB001580.

Helmstetter, A. and D. Sornette (2003a), Foreshocks explained by
cascades of triggered seismicity, J. Geophys. Res. (Solid Earth) 108
(B10), 2457 10.1029/2003JB002409 01.

Helmstetter, A. and D. Sornette (2003b) Bath's law Derived from the
Gutenberg-Richter law and from Aftershock Properties, Geophys. Res.
Lett., 30, 2069,  10.1029/2003GL018186.

Helmstetter, A. and D. Sornette (2003c) Importance of direct and
indirect triggered seismicity in the ETAS model of seismicity,
Geophys. Res. Lett. 30 (11) doi:10.1029/2003GL017670.

Helmstetter, A., D. Sornette and J.-R. Grasso (2003), Mainshocks are
Aftershocks of Conditional Foreshocks: How do foreshock statistical
properties emerge from aftershock laws, J. Geophys. Res., 108 (B10),
2046, doi:10.1029/2002JB001991.

Jones, L. M., R. Console, F. Di Luccio and M. Murru (1999) Are
foreshocks mainshocks whose aftershocks happen to be big? preprint
available at
http://pasadena.wr.usgs.gov/office/jones/italy-bssa.html

Kagan, Y.Y. and L. Knopoff (1980), Spatial distribution of
earthquakes: The two-point correlation function, Geophys. J. Roy.
Astr. Soc., 62, 303-320.

Kagan, Y.Y. and L. Knopoff (1981) Stochastic synthesis of earthquake
catalogs, J. Geophys. Res., 86, 2853-2862

Knopoff, L., Y.Y. Kagan and R. Knopoff (1982),
earthquakes sequences, Bull. Seism. Soc. Am. 72, 1663-1676.

Kossobokov, V.G. and S.A. Mazhkenov (1988), Spatial characteristics
of similarity for earthquake sequences: Fractality of seismicity,
Lecture Notes of the Workshop on Global Geophysical Informatics with
Applications to Research in Earthquake Prediction and Reduction of
Seismic Risk (15 Nov.-16 Dec., 1988), ICTP, 1988, Trieste, 15 p.

Lee, M.W., D. Sornette and L. Knopoff (1999), Persistence and
Quiescence of Seismicity on Fault Systems, Phys. Rev. Lett. 83,
4219-4222.

Lindman, M., K. Jonsdottir, R. Roberts, B. Lund, and R. B\"odvarsson
(2005), Earthquakes descaled: On waiting time distributions and
scaling laws, Phys. Rev. Lett. 94, 108501.

Livina, V.N., S. Havlin, and A. Bunde (2005), Memory in the
occurrence of earthquakes, Phys. Rev. Lett. 95, 208501.

Metzler, R. and J. Klafter (2000), The random walk's guide to
anomalous diffusion: a fractional dynamics approach, Physics Reports
339, 1-77

Molchan, G.M. (2005), Interevent time distribution of seismicity: a
theoretical approach, Pure appl. geophys., 162, 1135-1150.

Molchan, G. and T. Kronrod (2005), Seismic Interevent Time: A
Spatial Scaling and Multifractality, preprint physics/0512264
(2005).

Ogata, Y. (1988), Statistical models for earthquake occurrences and
residual analysis for point processes, J. Am. Stat. Assn. 83, 9-27.

Ogata, Y. (2005), Detection of anomalous seismicity as a  stress
change sensor, J. Geophys. Res., Vol.110, No.B5, B05S06,
doi:10.1029/2004JB003245.

Ogata Y. and Zhuang J. (2006), Space--time ETAS models and an
improved extension, Tectonophysics. 413 (1-2), 13-23.

Ouillon, G., C. Castaing and D. Sornette (1996), Hierarchical
scaling of faulting, J. Geophys. Res. 101, B3, 5477-5487.

Reasenberg, P. A. and Jones, L. M. (1989), Earthquake hazard after a
mainshock in California, Science 243, 1173-1176.

Reasenberg, P. A. and Jones, L. M. (1994), Earthquake aftershocks:
Update, Science 265, 1251-1252 (1994)

Saichev, A. and D. Sornette (2005), Distribution of the Largest
Aftershocks in Branching Models of Triggered Seismicity: Theory of
the Universal Bath's law, Phys. Rev. E 71, 056127.

Saichev, A. and D. Sornette (2006a), Power law distribution of
seismic rates: theory and data, Eur. J. Phys. B 49, 377-401.

Saichev, A. and D. Sornette (2006b), Renormalization of the ETAS
branching model of triggered seismicity from total to observable
seismicity, in press in Eur. Phys. J. B
 (http://arxiv.org/abs/physics/0507024)

Saichev, A. and D. Sornette (2006c) ``Universal'' Distribution of
Inter-Earthquake Times Explained, submitted to Phys. Rev. Letts.
(http://arxiv.org/abs/physics/0604018)

Schwartz, D. P. and K. J. Coppersmith (1984), Fault behavior and
characteristic earthquakes: Examples from the Wasatch and San
Andreas fault zones,  J. Geophys. Res., 89, 5681-5698.

Sieh, K.E. (1981), A review of geological evidence for recurrence
times of large earthquakes, in Earthquake Prediction -- An
International Review, Maurice Ewing Series 4 (the American
Geophysical Union).

Sornette, A. and D. Sornette (1999), Renormalization of earthquake
aftershocks, Geophys. Res. Lett. 6, N13, 1981-1984.

Sornette, D. (2004), Critical Phenomena in Natural Sciences, Chaos,
Fractals, Self-organization and Disorder: Concepts and Tools, 2nd
ed. (Springer Series in Synergetics, Heidelberg).

Sornette, D. and M.J. Werner (2005a), Constraints on the Size of the
Smallest Triggering Earthquake from the ETAS Model, Baath's Law, and
Observed Aftershock Sequences, J. Geophys. Res. 110, No. B8, B08304,
doi:10.1029/2004JB003535.

Sornette, D. and M.J. Werner (2005b), Apparent Clustering and
Apparent Background Earthquakes Biased by Undetected Seismicity, J.
Geophys. Res., Vol. 110, No. B9, B09303, 10.1029/2005JB003621,

Steacy, S., J. Gomberg and M. Cocco (2005), Introduction to special
section: Stress transfer, earthquake triggering, and time-dependent
seismic hazard, J. Geophys. Res. 110, B05S01,
doi:10.1029/2005JB003692.

Utsu, T., Y. Ogata and S. Matsu'ura (1995), The centenary of the
Omori Formula for a decay law of aftershock activity, J. Phys. Earth
43, 1-33.

Working Group on California Earthquake Probabilities (2003)
Earthquake Probabilities in the San Francisco Bay Region: 2002-2031,
Open-File Report 03-214, USGS.

Wyss, M., D. Schorlemmer, and S. Wiemer (2000), Mapping asperities
by minima of local recurrence time: San Jacinto-Elsinore fault
zones, J. Geophys. Res. 105 (B4), 7829-7844.

Zhuang J., Chang C.-P., Ogata Y. and Chen Y.-I. (2005), A study on
the background and clustering seismicity in the Taiwan region by
using a point process model, Journal of Geophysical Research, 110,
B05S18, doi:10.1029/2004JB003157.

Zhuang J., Ogata Y. and Vere-Jones D. (2002), Stochastic
declustering of space-time earthquake occurrences, Journal of the
American Statistical Association, 97, 369-380.

Zhuang J., Ogata Y. and Vere-Jones D. (2004), Analyzing earthquake
clustering features by using stochastic reconstruction, Journal of
Geophysical Research, 109, No. B5, B05301, doi:10.1029/2003JB002879.

\clearpage

\clearpage

\clearpage

\clearpage

\clearpage

\clearpage

\clearpage


\clearpage

\clearpage

\clearpage

\clearpage

\clearpage

\clearpage

\end{document}